\documentclass[sn-apa,iicol]{sn-jnl}

\usepackage{url}
\usepackage{amsmath}
\usepackage{graphicx}
\usepackage{booktabs}
\usepackage{multirow}
\usepackage{tabularx}
\usepackage{balance}
\usepackage{booktabs}
\usepackage{placeins}
\usepackage{threeparttable}

\jyear{2024}
\usepackage{fancyhdr} % for arxiv
\fancypagestyle{firststyle}
{
   \fancyhf{}
   \fancyfoot[C]{\scriptsize{\textit{First Monday}, vol. 29, no. 6. This is the author's copy. The publisher's copy is available online via \url{https://doi.org/10.5210/fm.v29i6.13357}.}}
}

\begin{document}

\title{Mysterious and Manipulative Black Boxes: A Qualitative Analysis of
  Perceptions on Recommender Systems}

\author*[1]{\fnm{Jukka} \sur{Ruohonen}}\email{juk@mmmi.sdu.dk}
\affil*[1]{\orgdiv{ M{\ae}rsk Mc-Kinney M\o{}ller Institute},
  \orgname{University of Southern Denmark}, \orgaddress{\city{S\o{}nderborg},
    \postcode{DK-6400}, \country{Denmark}}}

\abstract{Recommender systems are used to provide relevant suggestions on
  various matters. Although these systems are a classical research topic,
  knowledge is still limited regarding the public opinion about these
  systems. Public opinion is also important because the systems are known to
  cause various problems. To this end, this paper presents a qualitative
  analysis of the perceptions of ordinary citizens, civil society groups,
  businesses, and others on recommender systems in Europe.  The dataset examined
  is based on the answers submitted to a consultation about the Digital Services
  Act (DSA) recently enacted in the European Union (EU). Therefore, not only
  does the paper contribute to the pressing question about regulating new
  technologies and online platforms, but it also reveals insights about the
  policy-making of the DSA. According to the qualitative results, Europeans have
  generally negative opinions about recommender systems and the quality of their
  recommendations. The systems are widely seen to violate privacy and other
  fundamental rights. According to many Europeans, these also cause various
  societal problems, including even threats to democracy. Furthermore, existing
  regulations in the EU are commonly seen to have failed due to a lack of proper
  enforcement. Numerous suggestions were made by the respondents to the
  consultation for improving the situation, but only a few of these ended up to
  the DSA.}


\maketitle

\section{Introduction}

\thispagestyle{firststyle} % for arxiv

Recommender systems are computational software solutions for providing
suggestions that are most relevant for a particular user. These suggestions vary
from an application domain to another; these may refer to recommendations about
what to purchase, what news to consume, what music to listen, and so forth and
so on. Besides being a classical research topic in computer science, recommender
systems have long been important for delivering relevant information from the
vast sources of the Internet. These are also important for companies and their
business intelligence, including their online advertising. When a purchase is
nowadays made, an accompanying advertisement typically follows on an online
platform.

However, recommender systems have long been scrutinized and criticized for their
various ethical lapses~\citep{Milano20}. Privacy is typically and inevitably
violated because particularly the newer systems are based personalization. In
addition, concerns have frequently been raised about the accuracy and quality of
recommendations, their fairness and accountability, and explainability and
transparency of the systems. Recently, these systems have been seen to also
cause different individual harms and lead to different societal threats,
including but not limited to those for fundamental rights and even~democracy.

To this end and other ends, the EU enacted the DSA, that is, Regulation (EU)
2022/2065, in 2022. The general enforcement date is set to 2024. This regulation
covers numerous distinct issues related to online platforms. Also recommender
systems are covered. Thus, this paper examines the perceptions of ordinary EU
citizens, civil society groups who typically represent them particularly on
matters related to rights and technology, businesses, and others about
recommender systems based on the open consultation that was held for the DSA.

The motivation for the paper as well as its contribution are two-fold. First,
there is only a little qualitative research on the public perceptions of
recommender systems. Although quantitative surveys have been conducted in recent
years, qualitative insights have been lacking. Therefore, the paper's
qualitative approach and results patch an important gap in existing research. As
will be shown, the qualitative observations reveal interesting insights about
what people and various stakeholders ``really think'' as compared to their
answers to some Likert-scales in surveys.

Second, the paper contributes to the vast, timely, politically hot, and pressing
research domain on the regulation of new technologies, whether artificial
intelligence or face recognition, and the so-called Big Tech companies behind
these. While it is beyond the scope of this paper to delver deeper into this
topic, a motivating point can be made with a relation to technology
ethics.

There has been an interesting recent debate among ethics scholars and ethicists
regarding their stances on regulation. Besides the issues with ethics washing and
ethics bashing~\citep{Bietti20}, some have recently argued that rather than
pursuing ``ethics-from-within'' and promoting self-regulation, technology
ethicists should take the regulatory power of governments into account as a
viable solution for solving foundational problems~\citep{Saetra22}. Others have
disagreed, arguing along the lines that political philosophy and politics are
not without their own problems; there are political elites who can be likened to
Big Tech companies, politicians are vote-seekers who do not understand business
realities, public authorities are not always enforcing laws justly, and so
forth~\citep{Chomanski21}. What is striking in this ethics debate is
the lack of commonsense connections to the foundations of social
sciences, including political science and law.

When foundations of liberal democracies, such as fundamental rights and
democracy itself, are under a threat by technologies and their use, law is
expected to intervene~\citep{Hildebrandt20}, regardless of what ethicists think
and say. While privacy and data protection violations are good examples
  on the threats to fundamental rights, a good and timely, but hardly the only,
example of a threat to democracy is online election interference by foreign or
domestic bad actors. Furthermore, in liberal democracies it is the people, not
ethicists, who possess the ultimate power to decide over the laws they perceive
as necessary. This cornerstone applies even when keeping in mind the necessary
delegation of power in representative democracies and the EU's enduring
democracy deficit. With this point in mind, the paper's qualitative observations
about opinions and perceptions offer a lot to ponder also for ethicists.

For everyone, a central tenet throughout the paper is a hypothetical skepticism
of particularly European citizens and civil society groups toward recommendation
systems. Such skepticism was already present during the negotiations of the
General Data Protection Regulation (GDPR). Later on, signs of skepticism, which
to some extent and implicitly correlate with so-called
neo-luddism~\citep{Humberstone23}, have been present during the negotiations of
the EU's recent technology regulations, including the DSA. However, in politics
skepticism and critical viewpoints are one thing and business and other
realities another. Thus, the central tenet must be evaluated against other
interests and the outcome from the policy-making, the actual DSA. On these
notes, the paper's background should be better elaborated.

\section{Background}

The existing research on recommender systems is vast in computer
science. Recently, relevant contributions have been made also in numerous other
fields, including social sciences. Therefore, it suffices to only briefly skim
through a few relevant studies about public opinion and perceptions; further
relevant studies are pointed out during the presentation of the results. Thus,
to begin with, several empirical studies have recently been conducted about the
perceptions of people on algorithms and algorithmic decision-making.

The underlying questions are typically framed around fairness and
trust. According to literature reviews, however, there is no consensus over the
definitions for these two concepts, measurements on these vary from a study to
another, and results are generally ambiguous, indicating, for instance, that
humans are viewed more fairer compared to algorithms or the other way
around~\citep{Starke22, Treyger23}. Although it remains unclear whether fairness
can be even formalized mathematically~\citep{Buyl24}, some studies indicate a
middle-ground, finding support for an observation that algorithmic
decision-making and human-based decisions are both perceived as equally fair and
trustworthy in mechanical tasks~\citep{Lee18}. Regarding artificial intelligence
more generally, such factors as accountability, fairness, security, privacy,
accuracy, and explainability have been observed to matter regarding people's
perceptions~\citep{Kieslich22, Treyger23}. Analogous studies have been conducted
in the context of recommender systems.

A number of distinct dimensions and variables have been considered for
evaluating recommender systems. These include at least the following: the
perceived variety in recommendations, the accuracy and quality of
recommendations, the effort required to use a given system, the perceived
effectiveness, efficiency, and enjoyment, the difficulty in making choices based
on the recommendations, trust placed upon the systems and skepticism expressed
toward these, the availability of functionality for users to contribute to the
rankings and recommendations, scrutability of the systems, user interface
designs for the systems, compliance of the systems with regulations,
counterfactual recommendations, domain knowledge, and privacy
concerns~\citep{Knijnenburg11, Martijn22, Pu12, Shang22}. As recommender systems
are widely perceived as black boxes by people, particular emphasis has been
placed upon controllability (the ability for users to contribute) and
explainability, the latter focusing on making the recommendation process and the
reasons behind particular recommendations clearer~\citep{Tsai21}. Transparency
has often been seen as the primary way to achieve explainability or at least
improve it. Transparency also received a specific focus in the DSA.

Several surveys have been conduced with transparency in mind. Although the
causal presumptions in these survey studies are often ambiguous with little
uniformity across studies, transparency has been observed to improve people's
perceptions on the privacy and fairness of algorithmic decision-making in
general~\citep{Aysolmaza23}. Transparency of recommender systems also impacts
user satisfaction~\citep{Gedikli14}. It further fosters trust placed upon the
systems~\citep{Shin22b}. Trust, in turn, moderates privacy
concerns~\citep{Shin22a}. There are some notable weaknesses in these
studies. For instance, the underlying assumption seems to be that privacy is
merely a \textit{concern} of people; that privacy \textit{violations} would
somehow disappear with improved transparency. In other words, perceptions and
people's opinions do not necessarily match reality; privacy may be severely
violated even in case people have no concerns over it. This kind of reasoning is
present also in online advertisement research. Among other things, it has been
argued that people' privacy perceptions are malleable, which allows advertisers
to use different tactics and tricks to counter privacy
concerns~\citep{vanDenBroeck20}. Such reasoning is not followed in computer
science research on privacy. Nor is it the logic of data protection law.

Two additional brief points are warranted. First, as argued by Lessig in a
recent interview, recommender systems and generative artificial intelligence are
largely the same thing at the moment; they will be or already are
``\textit{deployed hand in glove in order to achieve the objective of the person
  deploying them}'' (\citealt{Patel23}; cf.~also \citealt{Kapoor23}). Note that
this particular person may not refer to a natural person. As will be shown, also
recommender systems, online advertisements, and privacy are closely---if not
inseparably---interlinked.  The second point follows: an important limitation in
existing research is about the harms caused by recommender systems and the
societal threats they pose. Although there is empirical research on the public
perception of threats caused by artificial intelligence~\citep{Kieslich21}, the
perceptions of citizens and other stakeholders about the societal threats caused
by recommender systems have received less attention. In fact, according to a
reasonable literature search, there are no directly comparable previous works on
this topic. Therefore, the paper fills an important gap in existing research. It
also contributes to the ongoing political debate around the DSA.

\section{Materials and Methods}

\subsection{Data}

The data examined is based on the responses to the DSA's open consultation
initiated by the \citet{EC20d}. The consultation period ran from June 2020 to
September 2020. In total, 2863 valid responses were received. The answers
solicited in the consultation covered the full range of the DSA: illegal and
harmful content, disinformation, systematic societal threats such as the
COVID-19 pandemic, content moderation and algorithms thereto, reporting
practices for illegal content, information sources for forensics, dispute
resolving for content takedowns and account suspensions, protection of minors,
platforms and marketplaces established outside of the EU, transparency reports
released by companies, data sharing between third-parties, disclosure of data to
competent authorities, threats to fundamental rights, platform liability,
specific questions about existing laws, unfair business practices of large
platforms and their gate-keeping roles, online advertising, media pluralism, and
so forth.

There were also two specific open-ended questions about recommender systems. The
answers given to these provide the empirical material for the examination. The
two questions are:

\begin{enumerate}

\itemsep 5pt

% This is column AY.
\item{\textit{``When content is recommended to you -- such as products to
    purchase on a platform, or videos to watch, articles to read, users to
    follow -- are you able to obtain enough information on why such content has
    been recommended to you?  Please explain.''}}

% This is column ES.
\item{\textit{``In your view, what measures are necessary with regard to algorithmic recommender systems used by online platforms?''}}

\end{enumerate}

The answers to these two questions are analyzed together; no attempts are made
to separate the answers and opinions on the availability of information from the
solution proposals. It should be also mentioned that many answers were given in
languages other than English. These were machine-translated with
Google~Translate. On that note, it should be further emphasized that the
  answers submitted to the open policy consultation are not bound to Europe. For
  instance, many non-European, globally operating technology companies and
  their lobby groups submitted their own answers.

\subsection{Methods}

Qualitative analysis is used for tackling the answers to the two broad and
predefined questions used in the survey about recommender systems. On one hand,
the answers are relatively short; typically only a few sentences were provided
by the respondents, although also longer answers with several paragraphs are
present. On the other hand, there are nearly three thousand answers in
total. This nature of the dataset limits the scope of suitable qualitative
methods; neither narratives nor discourses can be found from the answers, and so
forth. Thus, the analysis builds on three well-known methods: qualitative
content analysis, thematic analysis, and grounded theory.

Inductive logic is used for all three internally. However, externally, the three
methods were applied sequentially: grounded theory followed the thematic
analysis, which, in turn, followed the results from the qualitative content
analysis. This type of between-method triangulation is fairly common in
qualitative research~\citep{Flick04}. In the present context it puts the three
methods into a complementary relationship; the three methods complement each
others reciprocally.

In essence, a conventional variant of qualitative content analysis seeks to find
latent patterns and constructs by identifying key concepts and coding categories
from textual data during the analysis~\citep{Hsieh05}. As this method is prone
to reduce to mere counting~\citep{Morgan93}, which does not align well with the
rationale of qualitative analysis---given its goal of providing nuanced and
thick explanations, a thematic analysis was further used as a subsequent method.

The thematic method is highly similar to the qualitative content analysis
method. In essence: given the categories and counts already at hand, the answers
were re-read in order to specify the key themes characterizing the dataset
through clustering related answers together~\citep{Crowe15}. Finally, the
grounded theory variant adopted takes the key themes initially for granted, but
does not consider these as a theory, which requires transcending beyond plain
empirical descriptions~\citep{Apramian17}. Accordingly, theory requires a
refinement of the key themes such that an emerging theoretical core becomes
visible~\citep{Heath04}. As the analysis is about a regulation in the EU, it is
also the realm of European policy and politics to which the grounded theory
variant adopted seeks to transcend the key themes. Already because the DSA has
already been enacted, actual policy recommendations are still kept to a minimum,
but the transcended themes still contribute also to the practical policy
realm. In particular, the forthcoming enforcement of the new European regulation
is closely watched throughout the world.

In addition to the triangulation, the trustworthiness of the qualitative results
is improved by following the so-called principle of
transparency~\citep[cf.][]{Ruohonen20CHB}. That is, each notable qualitative
observation or claim is backed with an explicit reference to the dataset. These
numerical references refer to the rows in the (Excel) dataset. As the data is
openly available from the EU, this referencing allows also easy replication
checks.

\section{Results}

The results are presented according to the key themes obtained through the
thematic analysis, which, to recall, was conducted after the qualitative content
analysis. These key themes are: lack of information about recommender systems
(Section~\ref{subsec: black boxes}), privacy (Section~\ref{subsec: privacy}),
effectiveness of the recommendations (Section~\ref{subsec: recommendation
  effectiveness}), harms and threats caused by recommender systems
(Section~\ref{subsec: harms and threats}), proposals for countering some or all
of the harms and threats (Section \ref{subsec: solution proposals}), concerns
about the new regulation (Section~\ref{subsec: concerns}), and the DSA's already
ratified response~(Section \ref{subsec: dsa}). When possible and appropriate,
the presentation of these themes is accompanied with concise references to the
academic research literature. It should be also noted that the last theme in
Section~\ref{subsec: dsa} is not a part of the dataset; instead, it is a
critical reflection obtained through the grounded theory approach. It provides a
key reference point for the overall reflection in the concluding
Section~\ref{sec: conclusion}.

\subsection{Black Boxes}\label{subsec: black boxes}

Many---if not the majority---of individual EU citizens who responded to the
consultation expressed negative opinions about recommender systems. There are
many signs in the dataset about lack of knowledge, apathy, powerlessness, fear,
and anger toward algorithms and their recommendations.

Recommender systems are ``\textit{a complete black-box for me}''.\footnote{~543;
  cf. also 2006} Their recommendations were perceived as being mysterious by
many respondents.\footnote{~949; 1032; 1445; 2053; 2243; 2348} These just
appear.\footnote{~2066; 2320; 2421; 2852} They are just out
there.\footnote{~2150; 2529} It is baffling, very odd, and really
strange.\footnote{~2201; 2284; 2348} Therefore, people were suspicious about
recommender systems.\footnote{~889} Some mistrust the systems and oppose
algorithms.\footnote{~2359; 2439} The recommendations given by these were
annoying, surprising, and frightening to many.\footnote{~723; 741; 1186; 1399;
  2030; 2192} The recommendation process ``\textit{scares me a bit, to be
  honest}''.\footnote{~478} The systems also manipulate people and deceive the
public.\footnote{~1277; 1720; 2362; 2474} They are misleading, roach-like, and
unlawful.\footnote{~2813; 2826} Their recommendations are
cheeky\footnote{~2798}, unwanted\footnote{~2534}, obscure\footnote{~2336},
creepy\footnote{~2342; 2362}, dangerous\footnote{~2078},
unhealthy\footnote{~2457}, harmful\footnote{~1094; 2306},
inappropriate\footnote{~2139}, sexist\footnote{~2846}, and
offensive\footnote{~1543}. These recommendations further make it difficult to
disentangle true from false.\footnote{~1542} Some have fallen to scams because
of the recommendations.\footnote{~2296} The skepticism expressed does
  not stop here.

Many people did not want to be recommended by algorithms.\footnote{~662; 950;
  956; 1031; 1099; 1147; 1578} Nevertheless, recommender systems were still
forced upon them by coercive business practices.\footnote{~1475; 1712} These
systems pollute the Internet.\footnote{~1595} The never-ending ads blight
everything.\footnote{~2443} Such pollution makes it understandable why some
people were fundamentally opposed to individualized advertising, but these
people had no reasonable way for expressing their position.\footnote{~1405;
  2758} Some have still tried to suppress the recommendations.\footnote{~2189;
  2620} Some wanted more information so that they could block
these.\footnote{~689; cf. also 2146} At the same time, many companies thought
that consumers ``\textit{are interested in personalization}''.\footnote{~564}
The same assumption about personalization is often made in academic
research---and again without empirical backing or other proper
justifications~\citep{Shin22a}. These diverging opinions underline the
conflicting interests between platforms and their consumers or non-paying users.

There is a lack of information about recommender systems and the algorithms used
in these. Although some said that partial or full information was available
about recommender systems and their recommendations, the majority of respondents
expressed the opinion that there was no good information available~(see
Table~\ref{tab: information}). Even when some information is provided, it is
generally insufficient according to the respondents.

\begin{table}[th!b]
\centering
\caption{Availability of Information}
\label{tab: information}
\begin{tabular}{lc}
\toprule
Answer & Freq. \\
\hline
No sufficient information & 583 \\
Partially or fully sufficient information & 58 \\
\bottomrule
\end{tabular}
\end{table}

When citizens went to look for more information, they indeed often concluded
that the information provided was either too hard to understand or too general,
vague, ambiguous, misleading, and disguised.\footnote{~785; 964; 1327; 1503;
  1584; 1670; 1768; 1830; 1839; 1988; 1996; 2009; 2218; 2395; 2409; 2622; 2769;
  2782; 2792} These also lack translations to other languages than
English.\footnote{~2531} Some have tried to directly contact online advertisers
for information.\footnote{~859; 1091} But often ``\textit{it is impossible for
  users to trace them}''.\footnote{~262; cf. also 1182} Therefore, transparency
should be increased for allowing to identify who and which organizations are
pushing the recommendations.\footnote{~2638} Although access to advertising
profiles has been seen as important for data protection
rights~\citep{Hildebrandt09}, according to many respondents, it is impossible to
have a control over a profile generated for advertising.\footnote{~1074; 1126;
  1201; 1208; 2414; 2630}

Though, some have tried to exercise their data protection rights granted by the
GDPR. However, the results were often disappointing; ``\textit{data is never
  made available by companies displaying or producing recommendations, and is
  not available in GDPR requests}''.\footnote{~55; cf. also 840; 1190}
Analogously, a citizen expressed disappointment about both platforms and data
protection authorities after having dealt with a content deletion request with
two authorities in two different countries without success.\footnote{~826}
Furthermore, many people noted that the choices provided by some platforms did
not make a difference; they still received personalized advertisements even
though they had refused to give their consent.\footnote{~203; 740; 907; 1722;
  1141; 2238} Using the choices merely meant that the same amount of ads was
received but on slightly different topics.\footnote{~2379} In other words,
violations of the GDPR have been widespread and these likely continue today.

An important further point is that the ``why am I seeing this'' --type of
functionality offered by some platforms was widely seen as insufficient for
users, researchers, and public authorities to understand
recommendations.\footnote{~266; 983; 1001; 1020; 1403; 1419; 2097} Though,
unsurprisingly, a major industry lobby group stated the exact
opposite.\footnote{~1063} However, it is the critical viewpoint that receives
support from academic research; the transparency functionality of some platforms
has indeed been concluded to be insufficient, misleading, incomplete, and
vague~\citep{Andreou18}. The respondents also pointed out that the functionality
provided by other platforms is seldom actually used by people, possibly due to
the explicit reliance on deceptive user interface designs.\footnote{~1288; 1419}
Usability issues have been recognized also in recent
research~\citep{Armitage23}. This point was further raised by citizens;
information is difficult or even impossible to access and it is too
broad.\footnote{~1346; 2454; 2524; 2807} All in all, the apparent lack
  of transparency is a key element in the general skepticism expressed by the
  respondents.

\subsection{Tracking and Privacy}\label{subsec: privacy}

Privacy in the World Wide Web has continuously declined throughout the past
twenty years. Therefore, it is not surprising that tracking was also widely
acknowledged by the respondents; ``\textit{it seems obvious to me that
  algorithms are tracking me}''.\footnote{~11; cf. also 1523; 2421} Many felt
that they were under surveillance by algorithms of unknown entities who
harvested their personal data.\footnote{~2080} These entities rob people's
personal details.\footnote{~2225} The tracking was generally seen as intrusive
by many.\footnote{~739; 1247; 2353} In other words, it felt like
spying.\footnote{~1860; 2749} It was like ``\textit{espionage on a very large
  scale}''.\footnote{~721} Some people did not want to be visible to everyone in
the Internet.\footnote{~1095} But there was no way to escape.\footnote{~1207}
It was impossible to opt-out from tracking and receiving of
recommendations.\footnote{~1509; 1535; 1699} While people wanted to choose
whether to rely on algorithmic recommendations, the only option was to ignore
the suggestions.\footnote{~1703, 1727; 1961} There were no means to challenge
the recommendations.\footnote{~2250}

The general public also lacks knowledge.\footnote{~1235; 1712} Indeed, the
answers reveal differences in people's technical knowledge about tracking and
its countermeasures. Some merely said that it was all about
algorithms.\footnote{~2711; 2737} They know things.\footnote{~2282} They do
things.\footnote{~2834} Different platforms are somehow linked together but it
is impossible to understand how.\footnote{~2410}

For others, the starting point for trying to understand tracking was clear:
``\textit{I think it is related to cookies}''.\footnote{~319; also 685; 893;
  1120; 1810; 2088; 2345; 2350; 2353; 2561; 2685; 2778; 2784; 2793; 2794} In
fact, these ``\textit{cursed cookies are spread everywhere}''.\footnote{~628}
Some have tried to mitigate the situation by deleting cookies after each
browsing sessions and relying on ad-blockers and virtual private
networks.\footnote{~856; 1110; 1241; 1271; 1347; 2276; 2552; 2675} Even then,
some had to acknowledge that they still received personalized
advertisements.\footnote{~1246} Others acknowledged that ``\textit{accepting
  cookies is much faster than restricting or refusing cookies on most
  websites}''.\footnote{~617} Although some stated that the so-called cookie
banners contain sufficient information for understanding personalized
advertisements, it can be argued that the consent forms and banners brought by
the GDPR are neither usable nor working well.\footnote{~945} ``\textit{I~would
  have to switch off each service individually}'' so that ``\textit{it would
  take hours until I worked my way through to the switch-off/reject
  option}''.\footnote{~121; cf. also 1353} According to recent academic
  research, the opt-out functionality is not even working properly in most
  websites~\citep{LiuIqbal22}. Even if it would be and time would be taken to
switch the cookies off, websites would ask again in a few weeks.\footnote{~2551}

Besides cookies, many suspected that the recommendations were based on their
browsing histories and previous searches on search engines.\footnote{~666; 683;
  750; 891; 1088; 1190; 1283; 1287; 1373; 1407; 1553; 1807; 1884; 2096; 2146;
  2160; 2165; 2525; 2579; 2596; 2666; 2794; 2810} These are both correct
presumptions. Many suspected that recommendations were further based on their
past actions and the content consumed by their contacts and other users on the
same platform.\footnote{~675; 748; 1207; 1482; 1910; 1950; 2058; 2086; 2139;
  2143; 2154; 2487; 2491; 2685; 2741; 2778; 2789; 2818; 2847} Again, these
suspicions are correct.

Due to their reliance on intrusive but still often inaccurate tracking
techniques, recommender algorithms were seen as opaque and
inscrutable.\footnote{~1801; 1972; 2408; 2682} The opacity left plenty of room
for speculations in all directions.\footnote{~1691; 2008} As has been common
among social media users \citep{Eslami16}, such speculations were also presented
by some respondents. For instance, someone's recommendations were
``\textit{related to products my wife bought}''.\footnote{~949} Some other
speculated that there must be something intentional behind the recommendations
of conspiracy theories and hateful content after having previously consumed
content on social justice and the climate change.\footnote{~1911} Also another
respondent wondered why he or she was getting recommended climate denialism
after having watched videos on the climate change, concluding that platforms
engage in the dissemination of propaganda.\footnote{~2175} The same went for
viewing feminist content after which misogynist content soon started to follow
on a platform.\footnote{~2631} Others speculated further.

``\textit{Was a purchase made by my bank account shared from my bank with my
  eBook seller and fed into the recommendation?}'', asked one respondent who
further continued to speculate other possibilities that ``\textit{include the
  use of data from tracking cookies, tracking pixels and browser
  fingerprinting}''.\footnote{~61} Tracking pixels, third-party JavaScript code
and analytics, browser fingerprinting, and embedded videos and other content
were a concern raised also by many others.\footnote{~369; 556; 679; 1020; 1401;
  1419} All are also well-known and well-analyzed tracking techniques in
academic privacy research~\citep{Bekos23, Bermejo-Agueda23, Laperdrix20,
  Ruohonen17EISIC, Ruohonen18WPES, Ruohonen18PST}. But nowadays the privacy
infringing tracking techniques extend well-beyond the Web.

Among other things, algorithms ``\textit{sometimes also listen in on the
  mic}''.\footnote{~527} This listening presumption was shared also by many
others who felt that it is ``\textit{like my phone is secretly listening}'' or
that ``\textit{there is software listening to what I am
  saying}''.\footnote{~623; 2121; cf. also 2475} Smartphones ``\textit{hear}''
telephone conversions and ``\textit{pick up discussions}'', subsequently using
the picked information for advertisements.\footnote{~914; 1322; cf. also 2535}
Such listening is unacceptable and horrifying.\footnote{~2177; 1399} Given the
past privacy scandals and the ongoing practices with voice assistant
technologies used in smartphones and other products~\citep{Edu20, Iqbal23,
  Tabassum20}, it is difficult to objectively reassure such people that they are
just paranoid; that unauthorized algorithmic monitoring of phone calls for
advertising would be entirely out of the question.

Finally, two other well-known privacy-related topics are present in the
dataset. The first is the information asymmetry between technology companies and
users~\citep{Hjerppe23SOSYM}. This asymmetry is concisely summarized by the
following comment from a citizen; ``\textit{it is difficult for us but easy for
  them}''.\footnote{~1491} This asymmetry extends toward a wider information
inequality between platforms and everyone else, including citizens, researchers,
civil society, and public authorities.\footnote{~1403} To this end, some authors
have used a term epistemic inequality to describe the
situation~\citep{Zuboff20a}. Some citizens proposed a simple solution to the
inequality: consumers and people in general ``\textit{should have the same level
  of insight as the advertisers who selected the targets for their
  ads}''.\footnote{~58} The second topic is the famous privacy paradox: people
recognize that they are under surveillance but do little to protect
themselves~\citep{Barnes06}. In other words, the ``\textit{process of collecting
  this data and its use is not clear, which, however, does not worry the
  majority of Internet users, who without hesitation agree to the terms and
  conditions of various services}.''\footnote{~60} Though, from a European
perspective, it is not the duty of citizens alone to ensure that their
fundamental rights are respected. Regardless, privacy and data protection are
further key elements behind the overall skepticism expressed by the respondents
toward recommender systems.

\subsection{Recommendation Effectiveness}\label{subsec: recommendation effectiveness}

Some representatives from media companies stated that algorithms are good at
delivering relevant advertisements and promoting quality content.\footnote{~925}
Only a couple of citizens agreed, noting that recommender systems offered
exactly what was looked after.\footnote{~1848; 2765} Many others expressed the
opposite opinion; they were constantly bombarded with advertisements on products
and services they were not interested in.\footnote{~1071; 1090; 1492; 1518;
  2243; 2461; 2746} Also numerous other respondents felt that recommendations
and advertisements were often random, incorrect, and entirely unrelated to their
interests.\footnote{~1127; 1163; 1835; 1968; 2037; 2053; 2139; 2192; 2201; 2327;
  2368; 2597; 2706; 2860; 2861} This alleged lack of recommendation
  effectiveness may have contributed toward the skepticism.

An illuminating example would be employees working on alcohol policy and
alcohol-related treatment who were frequently targeted with content about
alcoholic beverages.\footnote{~187} As said, these are often also offensive to
people; before having died from cancer, someone's mother had received a
continuous stream of ads on miracle workers, coffins, and
burials.\footnote{~2798} As for the incorrect recommendations, someone said that
he or she is not interested in females but still received ads on mail-order
brides.\footnote{~635} Some other went to a jazz concert and was later
advertised insurance schemes.\footnote{~870} Such incorrect recommendations make
platforms look stupid according to some people.\footnote{~2814} Garbage in,
garbage out, as one respondent aptly summarized the situation for many
consumers.\footnote{~972}

Many respondents stated that they were immune to being influenced by recommender
systems.\footnote{~1105; 1140; 1205; 1252; 1275; 1529} According to some, people
can always use their own expertise for judgments and they always have the choice
on what to click.\footnote{~1424; 2672} Some only relied on recommendations if
these came from their own contacts or known third-parties.\footnote{~1238; 1326;
  2148; 2581; 2677; 2796} Advertisements were also seen as easily recognizable
and harmless as such.\footnote{~1166} These were a joke to some.\footnote{~2512}
Others simply did not care.\footnote{~1231} It was just
spamming.\footnote{~2297} These few opinions notwithstanding, some respondents
started from a presumption that recommender systems and their algorithms are
incredibly pervasive, possessing the capability to influence interests,
opinions, and behaviors, including social group formation.\footnote{~1288}
Indeed, there is a whole branch of academic literature on persuasion and
persuasive recommender systems that specifically seek to change people's
attitudes, behaviors, or both~\citep{Cremonesi12, Teppan15}. The basis for the
presumption becomes also evident when taking a look at the harms recommender
systems cause for individuals and the societal threats they nowadays
allegedly entail.

\subsection{Individual Harms and Societal Threats}\label{subsec: harms and threats}

In addition to the already realized threat to people's privacy and other
fundamental rights, numerous different harms and threats associated with
recommender systems were pointed out by the respondents. Both individual harms
and societal threats were raised. As can be seen from Table~\ref{tab: issues},
the most common concern was about disinformation, misinformation, and hate
speech.

\begin{table}[th!b]
\centering
\caption{Harms and Threats}
\label{tab: issues}
\begin{tabular}{lc}
\toprule
Category & Freq. \\
\hline
Disinformation, misinformation, and hate speech & 697 \\
Illegal, harmful, and offensive content or goods & 41 \\
Algorithmic biases and discrimination & 31 \\
Democracy, politics, elections, and polarization & 26 \\
Engagement, amplification, and emotionality & 21 \\
Competition, Big Tech, and market imperfections & 21 \\
Intellectual property and copyright infringements & 20 \\
Commercial promotion and priority rankings & 16 \\
Radicalization and extremism & 11 \\
Failure of the P2B regulation & 7 \\
Failure of the GDPR & 4 \\
Failure of the e-commerce directive & 2 \\
Consumerism and the climate change & 1 \\
\bottomrule
\end{tabular}
\end{table}

It should be mentioned that there were several identical answers in this
regard. These identical answers were submitted in English and particularly in
German and French. Although traceability is impossible without further
  data, it may thus be that a campaign was launched somewhere in the Internet
for delivering this particular message to the DSA's open
consultation. Nevertheless, there are still hundreds of unique and genuine
answers about disinformation, misinformation, propaganda, hate speech, and
related issues. The concern is real and pressing according to many EU citizens.

What started out as a simple idea about contacting with other people and sharing
photographs turned out to be a ``\textit{propaganda monster}''.\footnote{~2356}
It was really shocking to many that platforms are allowed to make money from
facilitating disinformation and hate.\footnote{~2476} It made people
mad.\footnote{~2495} The facilitation was seen as intolerable to any healthy
society.\footnote{~2206} The most common point in the hundreds of answers on
this topic was that amplification and monetization of these should be outlawed
and subjected to criminal justice. In other words, not only should those who
spread disinformation and hate be held responsible, but also platforms should be
subjected to both financial sanctions and criminal law as enablers of this
conduct.\footnote{~74; 1082; 1398; 1758; 1769; 2216, \textit{inter alia}} As
could be expected, this viewpoint of the majority was met with a concern about
censorship expressed by a minority.\footnote{~943; 1320; 1426; 1567; 1832; 2800}

The second most common concern was about illegal and harmful content, including
advertisement content for products and services. As there were numerous specific
questions on this topic in the consultation, including those related to content
moderation, it suffices here to only note a couple of points. The first is that
recommender systems were widely seen to promote illegal and harmful
content. According some, therefore, recommendations should not occur for
children and young people.\footnote{~909} The other point is that many
representatives of businesses, media, publishers, and cultural arts argued that
these systems should not recommend content that infringes copyrights and
violates intellectual property rights in general.

The third most common concern was about different biases that recommender
systems and their algorithms have. As this concern is well-recognized and
extensively studied in computer science, it again suffices to only note a few
general points. In general, these biases and the associated discrimination
involve those that the public have, those that are present in the training data,
and those that the individual developers of the recommender systems
have.\footnote{~241; 385; 1288; 2034; 2306; 2568; 2703} The consequences from
these biases vary; disinformation and hate, so-called filter bubbles and echo
chambers, radicalization, and related issues were commonly raised by the
respondents.\footnote{~129; 241; 484; 2082; 2381; 2407; 2503; 2688; 2835} On the
business side, biases were also seen to involve price
discrimination.\footnote{~287; 494} In a similar vein, streaming media were seen
as biased toward not recommending European and culturally diverse video
content.\footnote{~104} Platforms were also seen as biased in promoting free and
ad-funded content instead of subscription content.\footnote{~318} As could be
expected, they were also accused of being politically biased in their
actions.\footnote{~1213; 2516}

These various biases correlate with the fourth and fifth most common concerns
raised by the respondents: the effects of recommender systems upon politics,
democracy, and elections on one hand and their foundation upon engagement,
amplification, and emotionality on the other. In terms of the former, many
believed that societal fairness and democracy are under a threat because of the
platforms and their ``\textit{filth}''.\footnote{~2407; 2508} ``\textit{The
  whole system of making politics about sensations and emotions is a
  disaster}.''\footnote{~478} The Cambridge Analytica scandal and the mass
killings of the Rohingya minority in Myanmar were raised as alarming
precedents.\footnote{~296; 2487} The United States was also seen as a cautious
example of what recommender systems and algorithms can do to a
society.\footnote{~2381} According to these critical viewpoints, algorithms are
used by monopolists operating in a legal vacuum against people and their
rights.\footnote{~175; 2597}

Indeed, according to some, the public lacks knowledge about its own psychology;
therefore, it is dangerous to allow private salespeople and monopolist platforms
the power to manipulate societal ideologies.\footnote{~2610} Such manipulation
via recommender algorithms allows to control the masses.\footnote{~1675}. These
systems consequently lead to various abuses, ``\textit{the most dangerous being
  those related to the manipulation of the masses for political and electoral
  purposes, which represent a risk to democracy}.''\footnote{~64} What allows
these abuses is the business model of the platforms: the engagement that keeps
people addicted to the platforms through a continuous delivery of dopamine
doses, which, in turn, leads to the facilitation of disinformation, hate,
polarization, radicalization, extremism, and various other societal
ills.\footnote{~80; 186; 679; 1029} The current platforms and their recommender
systems are subject to ``\textit{fundamental rights abuse stemming from users'
  `engagements on steroids', economic revenue as an underlying reasoning behind
  open recommendations systems and dominant market position of these
  actors}.''\footnote{~373; cf. also 596; 679}

Hence, the abuses are closely tied to the dominant market position of Big Tech
companies. Regulating these companies ``\textit{is one of the biggest challenges
  we face globally today}.''\footnote{~2827} However, it is beyond the scope of
this paper to delve deeper into the various market imperfections that Big Tech
companies cause in Europe and elsewhere. Some examples can be still noted. For
instance, many respondents noted that platforms themselves prioritize their own
content and products unfairly.\footnote{~514; 659; 844; 1025; 1213} As is
well-known, media, publishers, and brands are deprived from their
advertisement revenues by platforms.\footnote{~318; 326} At the same time, as
was noted, citizens are deprived from their privacy and other fundamental
rights. Although the DSA is only a part of the EU's recent regulative efforts,
some people had already lost all hope: instead of attempting to solve the Big
Tech conundrum via regulation, the whole Internet should be rebuilt on the
principles of decentralization and open source according to
them.\footnote{~1213}

It is worth further stressing that many expressed an opinion that the EU's
previous regulative attempts have failed to deliver. In particular, businesses
get away from various data protection violations because the GDPR is not
properly enforced.\footnote{~2826} Nor does it impose any transparency
requirements.\footnote{~895; 2758} The many problems with the regulation's
enforcement are well-known also in academic research~\citep{Ruohonen22IS,
  Waldman21}. The recent decisions against Big Tech companies have done
  little to change their practices~\citep{Armitage23}. Another regulatory
  failure relates to Regulation (EU) 2019/1150, which is commonly known as the
  P2B regulation.

According to some, this regulation has done nothing to change the negotiation
imbalance between platforms and publishers, including the former's ability to
unilaterally dictate the terms.\footnote{~318} Even though the regulation
imposes some transparency requirements for the main parameters used to classify
and categorize content, it does not prevent the noted self-preference processing
through which platforms promote their own content.\footnote{~332} Nor has it
delivered in terms of fostering transparency in practice.\footnote{~102} The
regulation is also limited to relationships between platforms and businesses,
excluding the relationships between platforms and their users, which is relevant
in terms of content delivery.\footnote{~208} The problems with the outdated
e-commerce Directive 2000/31/EC are also well-recognized---in fact, these were
among the motivations of the European Commission for introducing the
DSA~\citep{Cauffman21, Heldt22}. To this end, some also suggested that the
e-commerce directive should be extended so that the directive's prohibition of
unsolicited email ads would become the norm for all online
advertising.\footnote{~859; 2758}

Finally, a brief reflection is required about the academic research on these
issues raised by the respondents. The existing research on Big Tech companies
and their engagement-based business model is extensive. The same applies for
research on disinformation. No references are required for delivering these
points; there are literally hundreds of relevant works on the these topics,
including several monographs.

However, there is no consensus over the societal effects of these; disagreements
exist regarding whether recommender systems and algorithmic solutions in general
cause political polarization, echo chambers and filter bubbles, hate,
radicalization, and related ills. According to recent studies, including
systematic literature reviews, there exists empirical evidence for the
underlying causal claims, but the evidence is still insufficient to draw
definite conclusions~(\citealt{Banaji22, CastanoPulgarin21, Ghayda18,
  Geissler23, Gowder23, Guess23, Iandoli21, Kubin21, Muhsin22, Ribeiro20,
  Smith22, Terren21, Tontodimmamma21, Whittaker21}; for non-academic articles
see, e.g., \citealt{Adams19, Ballard23, TTP23}). Recently, studies conducted on
behalf of the \citeauthor{EC23a} \citeyearpar{EC23a, EC23b} found that the
recommender systems of most large online platforms indeed amplify foreign
disinformation alongside terrorist, violent extremist, and borderline
content---and the existing countermeasures still seem
insufficient~\citep{EFCSN24}.  Such results have prompted some to again call
upon a suspension of recommender systems altogether \citep{ICCL23}.  If
platforms are unable to not recommend such content, it becomes moot to argue
that the problems would not be also technical, and that platforms could in
theory give users a choice to choose the content they wish to see
\citep[cf.~\textit{versus}][]{Kapoor23} Furthermore and alternatively,
triangulation with the results on the perception of generally poor
recommendation effectiveness~(see Section~\ref{subsec: recommendation
  effectiveness}) allows to also question the severity of the harms and threats
in practice. Be that as it may, it suffices at least to conclude that the
respondents' concerns are justified, given the nature of these as concise
political opinions delivered for a policy consultation. In addition, already the
number of distinct concerns allows to conclude that these have contributed to
the overall skepticism expressed by the consultation's respondents.

\subsection{Solution Proposals}\label{subsec: solution proposals}

Numerous different solutions were suggested by the respondents to the
consultation. These are summarized in Table~\ref{tab: solutions}. The most
obvious and common proposition is not listed: it is the transparency of
algorithms. While keeping this point in mind, the second most common solution
suggested was about choice; people should have a choice over whether they want
to use recommender systems.

\begin{table}[th!b]
\centering
\caption{Solutions Proposed}
\label{tab: solutions}
\begin{tabular}{lc}
\toprule
Category & Freq. \\
\hline
Possibility to opt-out or opt-in only by choice & 60 \\
Third-party audits, proofs, and verification & 50 \\
Education and explanations for laypeople & 43 \\
Prohibition of all (open) recommender systems & 37 \\
Human in the loop & 31 \\
Research, science, and civil society & 27 \\
Fundamental rights and European values & 19 \\
Enforcing the GDPR and the e-privacy directive & 17 \\
Full public disclosure of all algorithms & 15 \\
Promotion of media and journalism & 13 \\
Platform liability for content & 11 \\
Access to training data & 11 \\
Risk-analysis and impact assessments & 9 \\
Open source algorithms and free licenses & 7 \\
Alternative ranking criteria & 5 \\
Certifications for algorithms & 4 \\
Standardization of algorithms & 4 \\
Codes of conduct and self-regulation & 4 \\
Downgrading, fact-checking, and demonetization & 3 \\
Quality seals and badges & 2 \\
Values for algorithms & 1 \\
Content filters & 1 \\
\bottomrule
\end{tabular}
\end{table}

According to a less strict version of this choice proposal, people should have
an option to opt-out from automated recommender systems and use alternative
means for ranking. These ranking criteria include sorting chronologically,
alphabetically, according to price, and so forth.\footnote{~55, 1403} According
to a strict version of the proposal, these alternative ranking criteria should
be the default such that an explicit opt-in would be required for the reliance
on recommender system algorithms. If users would opt-in, they should have a
further right to object automatic recommendations.\footnote{~521; 1419} Children
and minors were again seen as a group who specifically should be given a
mandatory opt-out or an opt-in choice.\footnote{~820} In general, both
  the opt-out choice and the stricter opt-in choice reflect the foundational
  European concept of informational self-determination.

The second most common proposal was about third-party audits of recommender
systems and their algorithms. According to many, these should be conducted
either by competent public authorities or trusted academics.\footnote{~2154;
  2776} Competence and expertise were emphasized in this regard; ``\textit{it's
  a matter of professionals.}''\footnote{~655} Experts should be the ones asking
the questions.\footnote{~691; 2459} According to many respondents, such experts
should have also full access to the training data used for recommender
systems. Yet, as soon further clarified in the next section, these audits should
neither ``\textit{reveal insights into particular users}'' nor endanger trade
secrets.\footnote{~596; 1598} Because full public disclosure allows platform
manipulation, the audits should be only reserved for independent auditors bound
by secrecy, composed of technical experts with research
capabilities.\footnote{~697~2411} Especially politicians should steer away from
audits.\footnote{~2583} Despite such reservations of some respondents, a large
amount of citizens demanded that all recommender algorithms should be
``\textit{publicly available and transparent.}''\footnote{~589; cf. also 1189;
  1198; 1234; 1253; 1293; 1308; 1342; 2219; 2272; 2347} If such disclosure is
not possible or desirable, at least all inferred data should be made available
to users upon request according to some.\footnote{~656} The auditing proposal
was further accompanied with many ideas about practicalities. For instance,
there should be an ``\textit{algorithm officer}'' according to
some.\footnote{~1918; cf. also 388} Some others recommended institutionalized
data-sharing partnerships with open application programming
interfaces.\footnote{~475; 679; 2842} In line with the endorsements fro some
academics~\citep{Busch23}, it was also suggested that platforms should allow
third-parties to develop algorithms.\footnote{~578} Warrants for inspection were
noted in case platforms would refuse to~cooperate.\footnote{~194}

The third most common proposal related to education of people and different
explanations for recommender systems. Particularly the explanations have been
studied extensively in academic research for a long time~\citep{Gedikli14,
  Martijn22, McSherry05, Tintarev07}. Therefore, it suffices to only note a few
critical points raised by the respondents. Some argued that complicated
technical explanations are unlikely to be useful for users.\footnote{~464} At
the same time, like academics have done~\citep{Ruohonen23IPR}, others
argued that a mere summary of parameters used in machine learning models is not
sufficient.\footnote{~2842} It was also noted that many terms such as
explainability and interpretability are still vague---an argument shared
  also by many academics.\footnote{~493} Given such criticism, some recommended
different levels of explanations to different audiences; ``\textit{basic,
  applied, expert, academic.}''\footnote{~539} Such suggestions aligned with
arguments about a more general need for education; explanations for recommender
systems are not sufficient alone.\footnote{~1412} Educating citizens should
start already from the state education according to some.\footnote{~1162}
Finally, some business representatives argued that content recommendation
decisions are so complex that these would be difficult to explain to
users.\footnote{~2777} Given the intrusive but obscure global tracking
infrastructure that the companies have built over the years, it is no wonder
that they themselves feel incapable of explaining their recommendations and
advertisement choices.

The fourth most common proposal for solving the issues was simple: all
recommender systems should be simply prohibited. This drastic measure was
promoted by even a surprisingly large amount of EU citizens. ``\textit{Punish
  the algorithm!}''\footnote{~648} Ban recommender systems.\footnote{~502; 623;
  2232; 2526; 2672} Such short but decisive comments reflect the critical
attitude many people in Europe have toward recommender systems. Though, some
would be willing to have some concessions; some would only allow recommender
systems for goods and services.\footnote{~1390} Analogously, according to many
respondents, particularly recommending political content should be prohibited
together with medical advertising.\footnote{~662; 1020; 2148; 2319} In addition
to the many societal threats and individual harms, privacy and data protection
were often the rationale behind these prohibition
proposals. ``\textit{Fundamental ban}'' because the systems ``\textit{cannot be
  legally represented in compliance with the applicable requirements for the
  protection of personal data.}''\footnote{~549} In this regard, some urged that
the GDPR's Article~58(2) should be immediately invoked to impose a
ban.\footnote{~373} In addition to these prohibition calls, a large amount of
citizens argued that the problems would be easily solved if only the GDPR would
be strictly enforced together with the e-privacy Directive 2002/58/EC, which
regulates web cookies, among other things. This alleged failure of the
  previous regulations may have contributed to the critical viewpoints and the
  skepticism expressed.

Regarding the GDPR, there was also some confusion in the answers about whether
the statistical correlations and inferred data used in recommender systems are
personal data. According to some, these are not personal data and thus
enforcement via the GDPR cannot be done.\footnote{~266} However, some others
were sure that behavioral and inferred data fall under the GDPR.\footnote{~1403;
  1461} Given the GDPR's wording about personal data as any information relating
to an identified or identifiable person, existing interpretations including
those related to training data~\citep{Veale18}, and the fact that personalized
recommendations and advertisements explicitly target individual data
subjects based on their personal data, the verdict is on the side of the latter
arguments. In other words, the GDPR clearly applies.

These comments aligned with answers that emphasized human rights, fundamental
rights in the EU, and European values. Algorithms should generally
``\textit{stay within the boundaries of the Charter of Fundamental
  Rights}''.\footnote{~68; also 686; 1403; 1461} They should also promote
cultural diversity that ``\textit{is one of the pillars of the European Union’s
  founding texts}''.\footnote{~205; also 218; 974} Regarding such diversity,
recommender systems should particularly promote media pluralism in Europe,
public broadcasting, and European content in video streaming
services.\footnote{~129; 205; 318; 359; 427; 501; 974} In general, these systems
should obey data protection, rule of law, justice, proportionality, and
humanistic values.\footnote{~1149; cf. also 1158} To ensure compliance with
existing laws and fundamental rights, strict enforcement and harsh financial
penalties were recommended.\footnote{~255; 475; 653; 697} Any financial
incentives behind recommender systems should be disclosed or
removed.\footnote{~2831; 2173} Until platforms stop causing harms, their profits
should be depleted.\footnote{~2206} Platforms should also pay
their~taxes.\footnote{~2421; 2811} The overall skepticism is again highly
visible.

Human supervision of recommender systems was also a popular proposal. Despite a
growing interest also in academic research, many questions are still unclear
about human oversight for algorithms, including recommender
systems~\citep{Andersen23, Lai21}. When considering the size of the Big Tech
platforms, it also remains unclear how such human oversight would work in
practice particularly regarding harmful content and content~removals.

As can be concluded from Table~\ref{tab: solutions}, numerous other proposals
were also presented in the open consultation. Of these, liability deserves a
mention already because on the other side of the Atlantic, the debate has
largely been about the Communications Decency Act of 1996 and its shielding
provisions for platforms from legal liability regarding the content posted by
users of the platforms~\citep{Epstein20, Pagano18}. In this regard, the European
opinions differed to some extent. A larger group of respondents argued that
platforms should not have a get-out-of-jail card on content; they should be
treated as publishers.\footnote{~2113; 2206; 2321; 2395} By implication, they
should be subject to legal liability over content.\footnote{~339; 427; 2375;
  cf. also 807} In particular, some argued that platforms should not benefit
from the liability exemption specified in Article~14 of the e-commerce
directive.\footnote{~125} Disinformation and hate speech were seen as a specific
type of content to which liability should particularly apply.\footnote{~1119;
  2321} According to a minority group, however, platforms should not be held
liable for algorithmic flaws in recommender systems because these only incur a
low risk to users.\footnote{~353} Therefore, ``\textit{companies should benefit
  from broad immunity from liability for the recommendations or suggestions made
  by their algorithms.}''\footnote{~359} Many further points were also raised
regarding the potentially harmful consequences of the DSA for~businesses.

\subsection{Concerns}\label{subsec: concerns}

Business representatives responding to the consultation raised various distinct
concerns, which more or less conflicted with those expressed by citizens. A few
brief points are warranted about these concerns, which are summarized in
Table~\ref{tab: benefits}. To begin with, the usual neoliberal or libertarian
viewpoint is visible in the dataset; ``\textit{no regulation is
  needed}''.\footnote{~532; also 724; 997; 2823}. Algorithms should remain free
from any interference by governments, political parties, and non-governmental
organizations.\footnote{~1825; 1852} These viewpoints were accompanied with
concerns about competitiveness of smaller European companies;
``\textit{recommender systems are crucial for European scaleups to grow and
  compete}''.\footnote{~323}

\begin{table}[th!b]
\centering
\caption{Benefits, Non-Issues, and Concerns}
\label{tab: benefits}
\begin{tabular}{lc}
\toprule
Category & Freq. \\
\hline
Protection of trade secrets & 20 \\
Already addressed via the P2B regulation & 18 \\
Usefulness to users & 14 \\
Bad actors can exploit transparency & 13 \\
Right to appeal on content decisions & 6 \\
Already addressed via the GDPR & 6 \\
Already addressed via the omnibus directive & 5 \\
Transparency fosters trust & 4 \\
Freedom of expression must be ensured & 4 \\
Risks of hacking & 3 \\
Barriers to innovation & 2 \\
Editorial freedom must be ensured & 2 \\
Platforms promote freedom of expression & 2 \\
Liability threats & 2 \\
Bureaucracy and costs & 1 \\
% Already addressed via Directive 2010/13/EU & 1 \\
\bottomrule
\end{tabular}
\end{table}

Besides the antagonism toward all regulative action, the main concern of
businesses was about trade secrets that transparency requirements for algorithms
might reveal, as has been pointed out also by
academics~\citep{Turillazzi23}. Analogously to arguments of many
academics~\citep[p.~33]{Gowder23, Laufer23, Epstein20}, likewise, concerns were
also expressed that algorithmic transparency might expose systems to hacking and
that bad actors could exploit transparency to manipulate recommender
systems. Click and stream farms were used as examples about such
manipulation.\footnote{~133} To this end, Big Tech companies argued that even a
little amount of transparency endangers their trade secrets as well as the
security and integrity of their platforms and infrastructures, potentially
causing more harm than help to consumers and citizens.\footnote{~1598} Though,
interestingly enough, these critical points did not explicitly address the most
common (see Table~\ref{tab: solutions}) proposal for a solution; the possibility
to opt-out or opt-in only by choice.

As for trade secrets, a common point was that transparency was already addressed
in the so-called omnibus Directive 2019/2161 and particularly the P2B
regulation, which also provides legal guards against unwarranted disclosure of
technical details that might be used to manipulate ranking algorithms and
automated filtering decisions.\footnote{~147; 277; 301; 308; 464; 997; 1001} As
was noted earlier, however, not all businesses agreed with this claim made about
the P2B regulation. A related point raised was that the GDPR already supposedly
addressed some concerns.\footnote{~613} In particular, the regulation's
Article~22 for the opt-out possibility from automated decision-making and
profiling was seen as sufficient for avoiding further
regulations.\footnote{~178; 201; 277; 308} However, like academics have done
\citep{deHert21}, critics noted that the Article's wording about legal effects
or other similarly significantly affecting effects prevents citizens the
possibility to opt-out in the context of recommender systems.\footnote{~266;
  895; 2758}

Finally, various other concerns were further raised but to a lesser extent in
terms of volume. For instance, a concern was raised about mandating the use of
explainable algorithms, which, according to a respondent's viewpoint, prevent
the use of more advanced algorithms.\footnote{~464} Other concerns included
those related to automated content filters for recommender systems, which were
supported by some citizens.\footnote{~2150} According to critics, no automated
\textit{ex~ante} controls for content should be forced upon
companies.\footnote{~951; 1061} To this end, some noted that only \textit{ex
  post} enforcement should be considered as \textit{ex ante} measures are
useless.\footnote{~493} In a similar vein, a concern was also raised that
authorities might be able to dictate what type of content social media companies
should recommend, which would be particularly problematic in those European
countries with non-independent authorities and weak rule of law
provisions.\footnote{~359} In other words, the freedom of expression must be
guaranteed. Although the relation between content moderation and editorial
freedom was a more pressing issue~\citep{Papaevangelou23}, some media
representatives were concerned also about potential effects upon the editorial
freedom to choose the rankings for media content.\footnote{~244}

\subsection{The DSA's Answer}\label{subsec: dsa}

It is necessary to take a brief final look at what the actual DSA says about
recommender systems and imposes upon these. Recitals 55, 70, 84, 88, 94, 96, and
102 set the overall scene. The discussion in these recitals include ranking
demotion and shadow banning, suspension of monetization and advertising for bad
actors, transparency, risk assessments for countering systematic infringement,
testing of recommender algorithms, bias mitigation and data protection measures
particularly with respect to vulnerable groups and the GDPR's category of
sensitive personal data, availability of data for auditors, and
standardization. However, the actual regulatory mandates for recommender systems
are weak and limited in their scope. Only three such mandates are imposed.

First, Article~27 specifies transparency requirements for recommender
systems. These are simple enough: the main parameters used in the systems should
be specified in plain and intelligible language. The information about the main
parameters should include explanations about the most significant ranking
criteria and the reasons for the relevance of the parameters. The information
provided should cover also any potential options provided for users for altering
the parameters. Thus, these transparency requirements resemble those specified
in Article~5 of the P2B regulation. Overall, these are weak and easily subjected
to criticism. As has been already pointed out~\citep{Arcila23, Busch23,
  Helberger21}, specifying few vague sentences about main parameters in terms
and services hardly qualifies as transparency that would educate and
empower~people.

Second, Article~34 mandates very large online platforms (VLOPs) and very large
online search engines (VLOSEs) to carry out risk assessments. These cover also
recommender systems. Such assessments should address questions such as the
manipulation potential of the systems and their role in the amplification of
illegal content. Article~35 continues with risk mitigation, which includes
testing of recommender systems and their algorithms. Then, Article~40 further
mandates that VLOPs and VLOSEs should disclose the design, logic, functioning,
and testing of their recommender systems to competent regulatory coordinators or
the European Commission in particular. According to Article~44, the Commission
is also set to develop voluntary European standards, including those related
to choice interfaces and information on the main parameters. All in all, the
common proposal in the consultation about auditing was taken into account in the
DSA.

Third, Article~38 mandates VLOPs and VLOSEs to provide at least one ranking
criterion that is not based on profiling, as defined in the GDPR's Article~4. In
other words, Big Tech companies should provide at least one easily accessible
option that goes beyond personalization. Hence, the most common proposal of EU
citizens for an opt-out choice was to some extent taken into account, although
the stronger opt-in version of the proposal was bypassed by the lawmakers. As
was noted in Section~\ref{subsec: solution proposals}, both choices can be seen
to reflect the European concept of informational self-determination, and,
therefore, these were presumably relatively easy to lobby. It is finally worth
remarking that Article 38 has also been enforced in a recent court
case~\citep{Kupiec23}, but it remains to be seen whether the article's
intervention will change the recommender systems landscape more generally.

\section{Conclusion}\label{sec: conclusion}

This paper presented a qualitative analysis of the perceptions on recommender
systems by European citizens, civil society groups, public authorities,
businesses, and others. The dataset examined was based on the answers submitted
to the DSA's open consultation in 2020. The following eight points summarize the
qualitative results obtained:

\begin{itemize}

  \itemsep 5pt

\item{Recommender systems are widely perceived by Europeans and European
  stakeholders as black boxes. This perception is accompanied with negative
  opinions toward the systems and algorithms in general. The distinct adjectives
  used to describe the recommendations are revealing; according to numerous
  respondents, these are manipulative, deceptive, misleading, offensive,
  harmful, and so forth. Many people do not want to be recommended. Because some
  of the answers resemble also the common folk theories about recommendations on
  social media~\citep{Eslami16}, more transparent user interface design might
  counter at least some of the negative perceptions held by Europeans, whether
  citizens or stakeholders.}

\item{There is a consensus among the respondents that sufficient information is
  lacking about recommender systems. The existing information provided by some
  platforms is seen as vague and overly broad. Access to recommendation and
  advertising profiles is not provided. These opinions are shared by citizens,
  civil society groups, and researchers alike. Transparency is lacking.}

\item{The functionality of existing recommender systems is commonly seen to
  interlink with online tracking. Privacy and commercial surveillance are still
  a concern of many Europeans and their stakeholders. As with the
  recommender systems themselves, the whole tracking (a.k.a.~online
  advertisement) industry lacks transparency and proper judicial
  oversight. Existing laws are not properly enforced to an extent hoped by many
  EU citizens and European stakeholders.}

\item{The effectiveness of recommender systems is widely questioned. The quality
  and accuracy of recommendations are generally poor according to many
  Europeans. Given the opaque and inscrutable nature of these systems,
  which partially stems from their reliance on intrusive tracking, many people
  feel that recommendations and related advertisements are random, incorrect,
  and unrelated to their~interests.}

\item{Recommender systems are widely seen to cause individual harms and lead to
  societal threats. Disinformation, misinformation, and hate speech are the most
  pressing concern of Europeans and European stakeholders; the
  amplification and monetization of these should be made illegal. Recommender
  systems are commonly also seen to promote illegal, harmful, and offensive
  content. Different biases and discrimination are a further concern of
  many. What is more, these systems also threaten democracy and facilitate
  different societal ills, including political polarization, radicalization, and
  extremism. The engagement-based and profit-seeking business model behind
  recommender systems offered by near monopolies are to some extent behind these
  ills according to many Europeans, including businesses.}

\item{The most frequently raised proposal for reducing the various harms is
  simple: there should either be an opt-out choice or an explicit opt-in for
  automated recommender systems. Audits by competent public authorities,
  scientists, and other professionals are also commonly seen as a potential
  solution. Education and explanations are needed too. Furthermore, enforcement
  of existing laws, including the GDPR in particular, would solve many problems.
  Fundamental rights should be respected by recommender systems. Given the harms
  and threats, there exists also some support among EU citizens and European
  stakeholders for drastic measures such as a universal ban of these systems.}

\item{The most common concern about increasing transparency through regulation
  is about potential exposure of trade secrets. Existing regulations are also
  seen as sufficient according to some businesses. Transparency is seen to also
  expose recommender systems to manipulation by bad actors. Implications for
  other fundamental rights are also a concern of many; these include the freedom
  of expression in particular.}

\item{Given the numerous concerns of EU citizens and associated stakeholders,
  whether in terms of privacy violations or societal threats, the DSA fails to
  deliver in terms of recommender systems. That said, the regulation's scope is
  wide; the main focus is on content moderation and related platform issues.}

\end{itemize}

The central tenet---the skepticism and critical attitudes hypothesized, is
well-represented and visible in the dataset. Given the close connection of many
recommender systems to the online advertising business, some of the skepticism
is understandable, especially given the enduring problems in the GDPR's
enforcement \citep{Ruohonen23IPR, Ruohonen22IS}. It remains to be seen whether
the many problems can be even satisfactorily addressed without addressing the
many privacy violations too. Having said that, the individual harms and societal
threats are generally more difficult to evaluate, although it is clear that
these did not explicitly enter into the DSA's recommender system
scope. Furthermore, some of the proposals, such as the promotion of media and
journalism, are perhaps better addressed in other regulations, such as the
forthcoming European Media Freedom Act (EMFA).  Then, particularly the calls to
ban all recommender systems are close to neo-luddism. Even when such calls are
bypassed, it can be argued that the DSA left many questions unanswered with
respect to recommender systems.

Some limitations should be also acknowledged. Although qualitative analysis
itself is often open to criticism about subjectivity and researcher bias, a more
important concern relates to the dataset used; it may be biased. As the dataset
is about responses to a policy consultation for a regulative initiative, it may
well be that the responses are biased toward those who are particularly
interested in European politics and the EU's policy-making for technology. This
potential bias applies to citizens, civil society groups, academics, and
businesses alike, each of which may have their own biases due to their different
political interests and policy goals. By implication, the dataset cannot be
considered as representative to describe the perceptions and attitudes of
Europeans as a whole. This non-representativeness may also explain a portion of
the overall skepticism. While a comprehensive European survey would be required
to patch the representativeness issue, an alternative promising path for further
research would be to classify the answers according to the types of
respondents. It would allow to better examine the heavy lobbying, among other
things.

Also other biases may be present, including those related to socio-economic
factors. For instance, younger people have typically less concerns about privacy
and commercial surveillance~\citep{Kalmus22}. A further potential source of
biases stems from the consultation's broad scope, which may have led the
respondents to consider mainly the recommender systems of Big Tech companies,
hence excluding or downplaying considerations and opinions about smaller
recommender systems used in European online marketplaces, media, and other
domains.

As for further research, the noted biases translate to an important question
about the EU's technology-related policy-making. Little is known about the
politics surrounding recommender systems and artificial intelligence in general.
Who is promoting and lobbying what and why? Regarding such important questions,
contributions from political science, international relations, and associated
policy sciences are generally scarce~\citep{Ruohonen23CLSR,
  Srivastava21}. Despite the usual prodigious lobbying~\citep{Bendiek23}, many
observers maintain that actual political struggles were still only modest during
the DSA's surprisingly fast negotiations~\citep{Papaevangelou23,
  Schlag23}. Therefore, a plausible hypothesis can be also presented for further
research: the lobbying from Big Tech and other companies was successful in
limiting the regulatory scope of recommender systems to only a few relatively
weak mandates.

There is also some room for policy-making criticism about the EU's recent
regulative efforts. It seems that the EU is eager to pursue many new
regulations, while, at the same time, the enforcement of the existing ones,
including the GDPR in particular, is facing many problems. A~similar concern
remains about the DSA's future enforcement. Having learned from the GDPR's
enforcement problems, the lawmakers largely centralized the administration and
enforcement of the DSA to the EU-level and further required funding for the
enforcement from Big Tech and other companies.

Despite these measures, critics have already questioned whether meaningful
accountability will be delivered in practice due to potential enforcement
obstacles, coordination problems between national public authorities, procedural
issues, problems in auditing, incoherence in terms of national adaptations,
concerns over the freedom of expression, partial outsourcing to private sector,
and other problems~~\citep{Barata23b, Cauffman21, DFRLab23a, CEPA23a, CDT23a,
  Strowel23, Turillazzi23, VanCleynenbreugel23, vanHoboken22}. Against this
backdrop, all the craze around artificial intelligence should not overshadow the
foundational fact: transparency and accountability requirements apply also to
policy-makers, regulators, and public~administrations. On the bright side,
however, many of the DSA's fundamental goals, including those for recommender
systems, were recently more or less adopted also by \citet{UNESCO23}, indicating
that the so-called Brussels effect may perhaps apply also to this new type of a
platform regulation.

\section*{Acknowledgements}

This research was funded by the Strategic Research Council at the Academy of Finland (grant number 327391).

%\balance
%\bibliographystyle{apalike}
%\bibliography{eu}

\begin{thebibliography}{}
\providecommand{\doi}[1]{\url{https://doi.org/#1}}
\bibcommenthead

\bibitem [\protect \citeauthoryear {%
Adams%
}{%
Adams%
}{%
{\protect \APACyear {2019}}%
}]{%
Adams19}
\APACinsertmetastar {%
Adams19}%
\begin{APACrefauthors}%
Adams, R.%
\end{APACrefauthors}%
\unskip\
\newblock
\APACrefYearMonthDay{2019}{}{}.
\newblock
\APACrefbtitle {{S}ocial {M}edia {U}rged to {T}ake '{M}oment to {R}eflect'
  {A}fter {G}irl's {D}eath.} {{S}ocial {M}edia {U}rged to {T}ake '{M}oment to
  {R}eflect' {A}fter {G}irl's {D}eath.}
\newblock
\APACrefnote{{T}he {G}uardian. {A}vailable online in March 2024:
  \url{https://www.theguardian.com/media/2019/jan/30/social-media-urged-to-take-moment-to-reflect-after-girls-death}}
\PrintBackRefs{\CurrentBib}

\bibitem [\protect \citeauthoryear {%
Andersen%
\ \BBA {} Maalej%
}{%
Andersen%
\ \BBA {} Maalej%
}{%
{\protect \APACyear {2023}}%
}]{%
Andersen23}
\APACinsertmetastar {%
Andersen23}%
\begin{APACrefauthors}%
Andersen, J.S.%
\BCBT {}\ \BBA {} Maalej, W.%
\end{APACrefauthors}%
\unskip\
\newblock
\APACrefYearMonthDay{2023}{}{}.
\newblock
\APACrefbtitle {{D}esign {P}atterns for {M}achine {L}earning {B}ased {S}ystems
  with {H}uman-in-the-{L}oop.} {{D}esign {P}atterns for {M}achine {L}earning
  {B}ased {S}ystems with {H}uman-in-the-{L}oop.}
\newblock
\APACrefnote{{F}orthcoming in {IEEE} {S}oftware. {A}vailable online in
  December: \url{https://arxiv.org/abs/2312.00582}}
\PrintBackRefs{\CurrentBib}

\bibitem [\protect \citeauthoryear {%
Andreou%
\ \protect \BOthers {.}}{%
Andreou%
\ \protect \BOthers {.}}{%
{\protect \APACyear {2018}}%
}]{%
Andreou18}
\APACinsertmetastar {%
Andreou18}%
\begin{APACrefauthors}%
Andreou, A.%
, Venkatadri, G.%
, Goga, O.%
, Gummadi, K.P.%
, Loiseau, P.%
\BCBL {} Mislove, A.%
\end{APACrefauthors}%
\unskip\
\newblock
\APACrefYearMonthDay{2018}{}{}.
\newblock
{\BBOQ}\APACrefatitle {{I}nvestigating {A}d {T}ransparency {M}echanisms in
  {S}ocial {M}edia: {A} {C}ase {S}tudy of {F}acebook’s {E}xplanations}
  {{I}nvestigating {A}d {T}ransparency {M}echanisms in {S}ocial {M}edia: {A}
  {C}ase {S}tudy of {F}acebook’s {E}xplanations}.{\BBCQ}
\newblock
 \APACrefbtitle {{P}roceedings of the {N}etwork and {D}istributed {S}ystem
  {S}ecurity {S}ymposium ({NDSS} 2018)} {{P}roceedings of the {N}etwork and
  {D}istributed {S}ystem {S}ecurity {S}ymposium ({NDSS} 2018)}\ (\BPGS\ 1--15).
\newblock
\APACaddressPublisher{San Diego}{Internet Society}.
\PrintBackRefs{\CurrentBib}

\bibitem [\protect \citeauthoryear {%
Apramian%
, Cristancho%
, Watling%
\BCBL {}\ \BBA {} Lingard%
}{%
Apramian%
\ \protect \BOthers {.}}{%
{\protect \APACyear {2017}}%
}]{%
Apramian17}
\APACinsertmetastar {%
Apramian17}%
\begin{APACrefauthors}%
Apramian, T.%
, Cristancho, S.%
, Watling, C.%
\BCBL {} Lingard, L.%
\end{APACrefauthors}%
\unskip\
\newblock
\APACrefYearMonthDay{2017}{}{}.
\newblock
{\BBOQ}\APACrefatitle {{(Re)}{G}rounding {G}rounded {T}heory: {A} {C}lose
  {R}eading of {T}heory in {F}our {S}chools} {{(Re)}{G}rounding {G}rounded
  {T}heory: {A} {C}lose {R}eading of {T}heory in {F}our {S}chools}.{\BBCQ}
\newblock
\APACjournalVolNumPages{Qualitative Research}{17}{4}{359--376}.
\newblock

\newblock

\PrintBackRefs{\CurrentBib}

\bibitem [\protect \citeauthoryear {%
Arcila%
\ \BBA {} Griffin%
}{%
Arcila%
\ \BBA {} Griffin%
}{%
{\protect \APACyear {2023}}%
}]{%
Arcila23}
\APACinsertmetastar {%
Arcila23}%
\begin{APACrefauthors}%
Arcila, B.B.%
\BCBT {}\ \BBA {} Griffin, R.%
\end{APACrefauthors}%
\unskip\
\newblock
\APACrefYearMonthDay{2023}{}{}.
\newblock
\APACrefbtitle {{S}ocial {M}edia {P}latforms and {C}hallenges for {D}emocracy,
  {R}ule of {L}aw and {F}undamental {R}ights.} {{S}ocial {M}edia {P}latforms
  and {C}hallenges for {D}emocracy, {R}ule of {L}aw and {F}undamental
  {R}ights.}
\newblock
\APACrefnote{{E}uropean {P}arliament, {C}ommittee on {C}ivil {L}iberties,
  {J}ustice and {H}ome {A}ffairs {(LIBE)}. {A}vailable online in April 2023:
  \url{https://op.europa.eu/en/publication-detail/-/publication/ba4aec3f-dcca-11ed-a05c-01aa75ed71a1/language-en}}
\PrintBackRefs{\CurrentBib}

\bibitem [\protect \citeauthoryear {%
Armitage%
, Botton%
, {Dejeu-Castang}%
\BCBL {}\ \BBA {} Lemoine%
}{%
Armitage%
\ \protect \BOthers {.}}{%
{\protect \APACyear {2023}}%
}]{%
Armitage23}
\APACinsertmetastar {%
Armitage23}%
\begin{APACrefauthors}%
Armitage, C.%
, Botton, N.%
, {Dejeu-Castang}, L.%
\BCBL {} Lemoine, L.%
\end{APACrefauthors}%
\unskip\
\newblock
\APACrefYearMonthDay{2023}{}{}.
\newblock
\APACrefbtitle {{S}tudy on the {I}mpact of {R}ecent {D}evelopments in {D}igital
  {A}dvertising on {P}rivacy, {P}ublishers and {A}dvertisers.} {{S}tudy on the
  {I}mpact of {R}ecent {D}evelopments in {D}igital {A}dvertising on {P}rivacy,
  {P}ublishers and {A}dvertisers.}
\newblock
\APACrefnote{{E}uropean {C}ommission, {D}irectorate-{G}eneral for
  {C}ommunications {N}etworks, {C}ontent and {T}echnology. {A}vailable online
  in March 2023:
  \url{https://op.europa.eu/en/publication-detail/-/publication/8b950a43-a141-11ed-b508-01aa75ed71a1/}}
\PrintBackRefs{\CurrentBib}

\bibitem [\protect \citeauthoryear {%
Aysolmaza%
, M\"uller%
\BCBL {}\ \BBA {} Meacham%
}{%
Aysolmaza%
\ \protect \BOthers {.}}{%
{\protect \APACyear {2023}}%
}]{%
Aysolmaza23}
\APACinsertmetastar {%
Aysolmaza23}%
\begin{APACrefauthors}%
Aysolmaza, B.%
, M\"uller, R.%
\BCBL {} Meacham, D.%
\end{APACrefauthors}%
\unskip\
\newblock
\APACrefYearMonthDay{2023}{}{}.
\newblock
{\BBOQ}\APACrefatitle {{T}he {P}ublic {P}erceptions of {A}lgorithmic
  {D}ecision-{M}aking {S}ystems: {R}esults {F}rom a {L}arge-{S}cale {S}urvey}
  {{T}he {P}ublic {P}erceptions of {A}lgorithmic {D}ecision-{M}aking {S}ystems:
  {R}esults {F}rom a {L}arge-{S}cale {S}urvey}.{\BBCQ}
\newblock
\APACjournalVolNumPages{Telematics and Informatics}{79}{}{101954}.
\newblock

\newblock

\PrintBackRefs{\CurrentBib}

\bibitem [\protect \citeauthoryear {%
Ballard%
}{%
Ballard%
}{%
{\protect \APACyear {2023}}%
}]{%
Ballard23}
\APACinsertmetastar {%
Ballard23}%
\begin{APACrefauthors}%
Ballard, C.%
\end{APACrefauthors}%
\unskip\
\newblock
\APACrefYearMonthDay{2023}{}{}.
\newblock
\APACrefbtitle {{P}erhaps {Y}ou{T}ube {F}ixed {I}ts {A}lgorithm. {I}t {D}id
  {N}ot {F}ix {I}ts {E}xtremism {P}roblem.} {{P}erhaps {Y}ou{T}ube {F}ixed
  {I}ts {A}lgorithm. {I}t {D}id {N}ot {F}ix {I}ts {E}xtremism {P}roblem.}
\newblock
\APACrefnote{{T}ech {P}olicy {P}ress. {A}vailable online in November:
  \url{https://techpolicy.press/perhaps-youtube-fixed-its-algorithm-it-did-not-fix-its-extremism-problem/}}
\PrintBackRefs{\CurrentBib}

\bibitem [\protect \citeauthoryear {%
Banaji%
\ \BBA {} Bhat%
}{%
Banaji%
\ \BBA {} Bhat%
}{%
{\protect \APACyear {2022}}%
}]{%
Banaji22}
\APACinsertmetastar {%
Banaji22}%
\begin{APACrefauthors}%
Banaji, S.%
\BCBT {}\ \BBA {} Bhat, R.%
\end{APACrefauthors}%
\unskip\
\newblock
\APACrefYear{2022}.
\newblock
\APACrefbtitle {{S}ocial {M}edia and {H}ate} {{S}ocial {M}edia and {H}ate}.
\newblock
\APACaddressPublisher{New York}{Routledge}.
\PrintBackRefs{\CurrentBib}

\bibitem [\protect \citeauthoryear {%
Barata%
\ \BBA {} {Calvet-Bademunt}%
}{%
Barata%
\ \BBA {} {Calvet-Bademunt}%
}{%
{\protect \APACyear {2023}}%
}]{%
Barata23b}
\APACinsertmetastar {%
Barata23b}%
\begin{APACrefauthors}%
Barata, J.%
\BCBT {}\ \BBA {} {Calvet-Bademunt}, J.%
\end{APACrefauthors}%
\unskip\
\newblock
\APACrefYearMonthDay{2023}{}{}.
\newblock
\APACrefbtitle {{T}he {E}uropean {C}ommission's {A}pproach to {DSA} {S}ystemic
  {R}isk {I}s {C}oncerning for {F}reedom of {E}xpression.} {{T}he {E}uropean
  {C}ommission's {A}pproach to {DSA} {S}ystemic {R}isk {I}s {C}oncerning for
  {F}reedom of {E}xpression.}
\newblock
\APACrefnote{{T}ech {P}olicy {P}ress. {A}vailable online in March 2024:
  \url{https://www.techpolicy.press/the-european-commissions-approach-to-dsa-systemic-risk-is-concerning-for-freedom-of-expression/}}
\PrintBackRefs{\CurrentBib}

\bibitem [\protect \citeauthoryear {%
Barnes%
}{%
Barnes%
}{%
{\protect \APACyear {2006}}%
}]{%
Barnes06}
\APACinsertmetastar {%
Barnes06}%
\begin{APACrefauthors}%
Barnes, S.B.%
\end{APACrefauthors}%
\unskip\
\newblock
\APACrefYearMonthDay{2006}{}{}.
\newblock
{\BBOQ}\APACrefatitle {{A} {P}rivacy {P}aradox: {S}ocial {N}etworking in the
  {U}nited {S}tates} {{A} {P}rivacy {P}aradox: {S}ocial {N}etworking in the
  {U}nited {S}tates}.{\BBCQ}
\newblock
\APACjournalVolNumPages{First Monday}{11}{9}{}.
\newblock

\newblock

\PrintBackRefs{\CurrentBib}

\bibitem [\protect \citeauthoryear {%
Bekos%
, Papadopoulos%
, Markatos%
\BCBL {}\ \BBA {} Kourtellis%
}{%
Bekos%
\ \protect \BOthers {.}}{%
{\protect \APACyear {2023}}%
}]{%
Bekos23}
\APACinsertmetastar {%
Bekos23}%
\begin{APACrefauthors}%
Bekos, P.%
, Papadopoulos, P.%
, Markatos, E.P.%
\BCBL {} Kourtellis, N.%
\end{APACrefauthors}%
\unskip\
\newblock
\APACrefYearMonthDay{2023}{}{}.
\newblock
\APACrefbtitle {{T}he {H}itchhiker's {G}uide to {F}acebook {W}eb {T}racking
  with {I}nvisible {P}ixels and {C}lick {IDs}.} {{T}he {H}itchhiker's {G}uide
  to {F}acebook {W}eb {T}racking with {I}nvisible {P}ixels and {C}lick {IDs}.}
\newblock
\APACrefnote{{A}rchived manuscript. Available online in March 2023:
  \url{https://arxiv.org/abs/2208.00710}}
\PrintBackRefs{\CurrentBib}

\bibitem [\protect \citeauthoryear {%
Bendiek%
\ \BBA {} Stuerzer%
}{%
Bendiek%
\ \BBA {} Stuerzer%
}{%
{\protect \APACyear {2023}}%
}]{%
Bendiek23}
\APACinsertmetastar {%
Bendiek23}%
\begin{APACrefauthors}%
Bendiek, A.%
\BCBT {}\ \BBA {} Stuerzer, I.%
\end{APACrefauthors}%
\unskip\
\newblock
\APACrefYearMonthDay{2023}{}{}.
\newblock
{\BBOQ}\APACrefatitle {{T}he {B}russels {E}ffect, {E}uropean {R}egulatory
  {P}ower and {P}olitical {C}apital: {E}vidence for {M}utually {R}einforcing
  {I}nternal and {E}xternal {D}imensions of the {B}russels {E}ffect from the
  {E}uropean {D}igital {P}olicy {D}ebate} {{T}he {B}russels {E}ffect,
  {E}uropean {R}egulatory {P}ower and {P}olitical {C}apital: {E}vidence for
  {M}utually {R}einforcing {I}nternal and {E}xternal {D}imensions of the
  {B}russels {E}ffect from the {E}uropean {D}igital {P}olicy {D}ebate}.{\BBCQ}
\newblock
\APACjournalVolNumPages{Digital Society}{2}{5}{1--25}.
\newblock

\newblock

\PrintBackRefs{\CurrentBib}

\bibitem [\protect \citeauthoryear {%
{Bermejo-Agueda}%
, Callejo%
, Cuevas%
\BCBL {}\ \BBA {} \'engel Cuevas%
}{%
{Bermejo-Agueda}%
\ \protect \BOthers {.}}{%
{\protect \APACyear {2023}}%
}]{%
Bermejo-Agueda23}
\APACinsertmetastar {%
Bermejo-Agueda23}%
\begin{APACrefauthors}%
{Bermejo-Agueda}, M.A.%
, Callejo, P.%
, Cuevas, R.%
\BCBL {} \'engel Cuevas.%
\end{APACrefauthors}%
\unskip\
\newblock
\APACrefYearMonthDay{2023}{}{}.
\newblock
{\BBOQ}\APACrefatitle {{adF}: {A} {N}ovel {S}ystem for {M}easuring {W}eb
  {F}ingerprinting through {A}ds} {{adF}: {A} {N}ovel {S}ystem for {M}easuring
  {W}eb {F}ingerprinting through {A}ds}.{\BBCQ}
\newblock

\newblock
\APACrefnote{{A}rchived manuscript. {A}vailable online in November:
  \url{https://arxiv.org/abs/2311.08769}}
\newblock

\newblock

\PrintBackRefs{\CurrentBib}

\bibitem [\protect \citeauthoryear {%
Bhatia%
\ \BBA {} Allen%
}{%
Bhatia%
\ \BBA {} Allen%
}{%
{\protect \APACyear {2023}}%
}]{%
CDT23a}
\APACinsertmetastar {%
CDT23a}%
\begin{APACrefauthors}%
Bhatia, A.%
\BCBT {}\ \BBA {} Allen, A.%
\end{APACrefauthors}%
\unskip\
\newblock
\APACrefYearMonthDay{2023}{}{}.
\newblock
\APACrefbtitle {{A}uditing in the {D}ark: {G}uidance is {N}eeded to {E}nsure
  {M}aximum {I}mpact of {DSA} {A}lgorithmic {A}udits.} {{A}uditing in the
  {D}ark: {G}uidance is {N}eeded to {E}nsure {M}aximum {I}mpact of {DSA}
  {A}lgorithmic {A}udits.}
\newblock
\APACrefnote{{C}enter for {D}emocracyu \& {T}echnology ({CDT}). {A}vailable
  online in December 2023:
  \url{https://cdt.org/insights/auditing-in-the-dark-guidance-is-needed-to-ensure-maximum-impact-of-dsa-algorithmic-audits/}}
\PrintBackRefs{\CurrentBib}

\bibitem [\protect \citeauthoryear {%
Bietti%
}{%
Bietti%
}{%
{\protect \APACyear {2020}}%
}]{%
Bietti20}
\APACinsertmetastar {%
Bietti20}%
\begin{APACrefauthors}%
Bietti, E.%
\end{APACrefauthors}%
\unskip\
\newblock
\APACrefYearMonthDay{2020}{}{}.
\newblock
{\BBOQ}\APACrefatitle {{F}rom {E}thics {W}ashing to {E}thics {B}ashing: {A}
  {V}iew on {T}ech {E}thics from {W}ithin {M}oral {P}hilosophy} {{F}rom
  {E}thics {W}ashing to {E}thics {B}ashing: {A} {V}iew on {T}ech {E}thics from
  {W}ithin {M}oral {P}hilosophy}.{\BBCQ}
\newblock
 \APACrefbtitle {{P}roceedings of the {C}onference on {F}airness,
  {A}ccountability, and {T}ransparency ({FAT* 2020)}} {{P}roceedings of the
  {C}onference on {F}airness, {A}ccountability, and {T}ransparency ({FAT*
  2020)}}\ (\BPGS\ 210--219).
\newblock
\APACaddressPublisher{Barcelona}{ACM}.
\PrintBackRefs{\CurrentBib}

\bibitem [\protect \citeauthoryear {%
Busch%
}{%
Busch%
}{%
{\protect \APACyear {2023}}%
}]{%
Busch23}
\APACinsertmetastar {%
Busch23}%
\begin{APACrefauthors}%
Busch, C.%
\end{APACrefauthors}%
\unskip\
\newblock
\APACrefYearMonthDay{2023}{}{}.
\newblock
{\BBOQ}\APACrefatitle {{F}rom {A}lgorithmic {T}ransparency to {A}lgorithmic
  {C}hoice: {E}uropean {P}erspectives on {R}ecommender {S}ystems and {P}latform
  {R}egulation} {{F}rom {A}lgorithmic {T}ransparency to {A}lgorithmic {C}hoice:
  {E}uropean {P}erspectives on {R}ecommender {S}ystems and {P}latform
  {R}egulation}.{\BBCQ}
\newblock
 S.~Genovesi, K.~Kaesling\BCBL {}\ \BBA {} S.~Robbins\ (\BEDS), \APACrefbtitle
  {{R}ecommender {S}ystems: {L}egal and {E}thical {I}ssues} {{R}ecommender
  {S}ystems: {L}egal and {E}thical {I}ssues}\ (\BPGS\ 31--54).
\newblock
\APACaddressPublisher{Cham}{Springer}.
\PrintBackRefs{\CurrentBib}

\bibitem [\protect \citeauthoryear {%
Buyl%
\ \BBA {} {De Bie}%
}{%
Buyl%
\ \BBA {} {De Bie}%
}{%
{\protect \APACyear {2024}}%
}]{%
Buyl24}
\APACinsertmetastar {%
Buyl24}%
\begin{APACrefauthors}%
Buyl, M.%
\BCBT {}\ \BBA {} {De Bie}, T.%
\end{APACrefauthors}%
\unskip\
\newblock
\APACrefYearMonthDay{2024}{}{}.
\newblock
{\BBOQ}\APACrefatitle {{I}nherent {L}imitations of {AI} {F}airness} {{I}nherent
  {L}imitations of {AI} {F}airness}.{\BBCQ}
\newblock
\APACjournalVolNumPages{Communications of the ACM}{67}{2}{48--55}.
\newblock

\newblock

\PrintBackRefs{\CurrentBib}

\bibitem [\protect \citeauthoryear {%
{Casta\~{n}o-Pulgar\'in}%
, {Su\'arez-Betancur}%
, Vega%
\BCBL {}\ \BBA {} L\'opez%
}{%
{Casta\~{n}o-Pulgar\'in}%
\ \protect \BOthers {.}}{%
{\protect \APACyear {2021}}%
}]{%
CastanoPulgarin21}
\APACinsertmetastar {%
CastanoPulgarin21}%
\begin{APACrefauthors}%
{Casta\~{n}o-Pulgar\'in}, S.A.%
, {Su\'arez-Betancur}, N.%
, Vega, L.M.T.%
\BCBL {} L\'opez, H.M.H.%
\end{APACrefauthors}%
\unskip\
\newblock
\APACrefYearMonthDay{2021}{}{}.
\newblock
{\BBOQ}\APACrefatitle {{I}nternet, {S}ocial {M}edia and {O}nline {H}ate
  {S}peech. {S}ystematic {R}eview} {{I}nternet, {S}ocial {M}edia and {O}nline
  {H}ate {S}peech. {S}ystematic {R}eview}.{\BBCQ}
\newblock
\APACjournalVolNumPages{Aggression and Violent Behavior}{58}{}{101608}.
\newblock

\newblock

\PrintBackRefs{\CurrentBib}

\bibitem [\protect \citeauthoryear {%
Cauffman%
\ \BBA {} Goanta%
}{%
Cauffman%
\ \BBA {} Goanta%
}{%
{\protect \APACyear {2021}}%
}]{%
Cauffman21}
\APACinsertmetastar {%
Cauffman21}%
\begin{APACrefauthors}%
Cauffman, C.%
\BCBT {}\ \BBA {} Goanta, C.%
\end{APACrefauthors}%
\unskip\
\newblock
\APACrefYearMonthDay{2021}{}{}.
\newblock
{\BBOQ}\APACrefatitle {{A} {N}ew {O}rder: {T}he {D}igital {S}ervices {A}ct and
  {C}onsumer {P}rotection} {{A} {N}ew {O}rder: {T}he {D}igital {S}ervices {A}ct
  and {C}onsumer {P}rotection}.{\BBCQ}
\newblock
\APACjournalVolNumPages{European Journal of Risk Regulation}{12}{}{758--774}.
\newblock

\newblock

\PrintBackRefs{\CurrentBib}

\bibitem [\protect \citeauthoryear {%
Chomanski%
}{%
Chomanski%
}{%
{\protect \APACyear {2021}}%
}]{%
Chomanski21}
\APACinsertmetastar {%
Chomanski21}%
\begin{APACrefauthors}%
Chomanski, B.%
\end{APACrefauthors}%
\unskip\
\newblock
\APACrefYearMonthDay{2021}{}{}.
\newblock
{\BBOQ}\APACrefatitle {{T}he {M}issing {I}ngredient in the {C}ase for
  {R}egulating {B}ig {T}ech} {{T}he {M}issing {I}ngredient in the {C}ase for
  {R}egulating {B}ig {T}ech}.{\BBCQ}
\newblock
\APACjournalVolNumPages{Minds and Machines}{31}{}{257--275}.
\newblock

\newblock

\PrintBackRefs{\CurrentBib}

\bibitem [\protect \citeauthoryear {%
Cremonesi%
, Garzotto%
\BCBL {}\ \BBA {} Turrin%
}{%
Cremonesi%
\ \protect \BOthers {.}}{%
{\protect \APACyear {2012}}%
}]{%
Cremonesi12}
\APACinsertmetastar {%
Cremonesi12}%
\begin{APACrefauthors}%
Cremonesi, P.%
, Garzotto, F.%
\BCBL {} Turrin, R.%
\end{APACrefauthors}%
\unskip\
\newblock
\APACrefYearMonthDay{2012}{}{}.
\newblock
{\BBOQ}\APACrefatitle {{I}nvestigating the {P}ersuasion {P}otential of
  {R}ecommender {S}ystems from a {Q}uality {P}erspective: {A}n {E}mpirical
  {S}tudy} {{I}nvestigating the {P}ersuasion {P}otential of {R}ecommender
  {S}ystems from a {Q}uality {P}erspective: {A}n {E}mpirical {S}tudy}.{\BBCQ}
\newblock
\APACjournalVolNumPages{ACM Transactions on Interactive Intelligent
  Systems}{2}{2}{1--41}.
\newblock

\newblock

\PrintBackRefs{\CurrentBib}

\bibitem [\protect \citeauthoryear {%
Crowe%
, Inder%
\BCBL {}\ \BBA {} Porter%
}{%
Crowe%
\ \protect \BOthers {.}}{%
{\protect \APACyear {2015}}%
}]{%
Crowe15}
\APACinsertmetastar {%
Crowe15}%
\begin{APACrefauthors}%
Crowe, M.%
, Inder, M.%
\BCBL {} Porter, R.%
\end{APACrefauthors}%
\unskip\
\newblock
\APACrefYearMonthDay{2015}{}{}.
\newblock
{\BBOQ}\APACrefatitle {{C}onducting {Q}ualitative {R}esearch in {M}ental
  {H}ealth: {T}hematic and {C}ontent {A}nalyses} {{C}onducting {Q}ualitative
  {R}esearch in {M}ental {H}ealth: {T}hematic and {C}ontent {A}nalyses}.{\BBCQ}
\newblock
\APACjournalVolNumPages{Australian \& New Zealand Journal of
  Psychiatry}{49}{7}{616-623}.
\newblock

\newblock

\PrintBackRefs{\CurrentBib}

\bibitem [\protect \citeauthoryear {%
{de Hert}%
\ \BBA {} Lazcoz%
}{%
{de Hert}%
\ \BBA {} Lazcoz%
}{%
{\protect \APACyear {2021}}%
}]{%
deHert21}
\APACinsertmetastar {%
deHert21}%
\begin{APACrefauthors}%
{de Hert}, P.%
\BCBT {}\ \BBA {} Lazcoz, G.%
\end{APACrefauthors}%
\unskip\
\newblock
\APACrefYearMonthDay{2021}{}{}.
\newblock
\APACrefbtitle {{R}adical {R}ewriting of {A}rticle 22 {GDPR} on {M}achine
  {D}ecisions in the {AI} {E}ra.} {{R}adical {R}ewriting of {A}rticle 22 {GDPR}
  on {M}achine {D}ecisions in the {AI} {E}ra.}
\newblock
\APACrefnote{{E}uropean {L}aw {B}log. {A}vailable online in {F}ebruary 2023:
  \url{https://europeanlawblog.eu/2021/10/13/radical-rewriting-of-article-22-gdpr-on-machine-decisions-in-the-ai-era/}}
\PrintBackRefs{\CurrentBib}

\bibitem [\protect \citeauthoryear {%
Edu%
, Such%
\BCBL {}\ \BBA {} {Suarez-Tangil}%
}{%
Edu%
\ \protect \BOthers {.}}{%
{\protect \APACyear {2020}}%
}]{%
Edu20}
\APACinsertmetastar {%
Edu20}%
\begin{APACrefauthors}%
Edu, J.S.%
, Such, J.M.%
\BCBL {} {Suarez-Tangil}, G.%
\end{APACrefauthors}%
\unskip\
\newblock
\APACrefYearMonthDay{2020}{}{}.
\newblock
{\BBOQ}\APACrefatitle {{S}mart {H}ome {P}ersonal {A}ssistants: {A} {S}ecurity
  and {P}rivacy {R}eview} {{S}mart {H}ome {P}ersonal {A}ssistants: {A}
  {S}ecurity and {P}rivacy {R}eview}.{\BBCQ}
\newblock
\APACjournalVolNumPages{ACM Computing Surveys}{53}{6}{1--36}.
\newblock

\newblock

\PrintBackRefs{\CurrentBib}

\bibitem [\protect \citeauthoryear {%
{EFCSN}%
}{%
{EFCSN}%
}{%
{\protect \APACyear {2024}}%
}]{%
EFCSN24}
\APACinsertmetastar {%
EFCSN24}%
\begin{APACrefauthors}%
{EFCSN}%
\end{APACrefauthors}%
\unskip\
\newblock
\APACrefYearMonthDay{2024}{}{}.
\newblock
\APACrefbtitle {{F}act-{C}hecking and {R}elated {R}isk-{M}itigation {M}easures
  for {D}isinformation in the {V}ery {L}arge {O}nline {P}latforms and {S}earch
  {E}ngines.} {{F}act-{C}hecking and {R}elated {R}isk-{M}itigation {M}easures
  for {D}isinformation in the {V}ery {L}arge {O}nline {P}latforms and {S}earch
  {E}ngines.}
\newblock
\APACrefnote{{E}uropean {F}act-{C}hecking {S}tandard {N}etwork ({EFCSN}).
  {A}vailable online in January:
  \url{https://efcsn.com/app/uploads/2024/01/FINAL_Fact_checking_and_related_Risk_Mitigation_Measures_for_Disinformation.pdf}}
\PrintBackRefs{\CurrentBib}

\bibitem [\protect \citeauthoryear {%
Epstein%
}{%
Epstein%
}{%
{\protect \APACyear {2020}}%
}]{%
Epstein20}
\APACinsertmetastar {%
Epstein20}%
\begin{APACrefauthors}%
Epstein, B.%
\end{APACrefauthors}%
\unskip\
\newblock
\APACrefYearMonthDay{2020}{}{}.
\newblock
{\BBOQ}\APACrefatitle {{W}hy {I}t {I}s {S}o {D}ifﬁcult to {R}egulate
  {D}isinformation {O}nline} {{W}hy {I}t {I}s {S}o {D}ifﬁcult to {R}egulate
  {D}isinformation {O}nline}.{\BBCQ}
\newblock
 W.L.~Bennett\ \BBA {} S.~Livingston\ (\BEDS), \APACrefbtitle {{T}he
  {D}isinformation {A}ge: {P}olitics, {T}echnology, and {D}isruptive
  {C}ommunication in the {U}nited {S}tates} {{T}he {D}isinformation {A}ge:
  {P}olitics, {T}echnology, and {D}isruptive {C}ommunication in the {U}nited
  {S}tates}\ (\BPGS\ 190--210).
\newblock
\APACaddressPublisher{Cambridge}{Cambridge University Press}.
\PrintBackRefs{\CurrentBib}

\bibitem [\protect \citeauthoryear {%
Eslami%
\ \protect \BOthers {.}}{%
Eslami%
\ \protect \BOthers {.}}{%
{\protect \APACyear {2016}}%
}]{%
Eslami16}
\APACinsertmetastar {%
Eslami16}%
\begin{APACrefauthors}%
Eslami, M.%
, Karahalios, K.%
, Sandvig, C.%
, Vaccaro, K.%
, Rickman, A.%
, Hamilton, K.%
\BCBL {} Kirlik, A.%
\end{APACrefauthors}%
\unskip\
\newblock
\APACrefYearMonthDay{2016}{}{}.
\newblock
{\BBOQ}\APACrefatitle {{F}irst {I} ''{L}ike'' {I}t, {T}hen {I} {H}ide {I}t:
  {F}olk {T}heories on {S}ocial {F}eeds} {{F}irst {I} ''{L}ike'' {I}t, {T}hen
  {I} {H}ide {I}t: {F}olk {T}heories on {S}ocial {F}eeds}.{\BBCQ}
\newblock
 \APACrefbtitle {{P}roceedings of the {ACM} {C}onference on {H}uman {F}actors
  in {C}omputing {S}ystems ({CHI 2016})} {{P}roceedings of the {ACM}
  {C}onference on {H}uman {F}actors in {C}omputing {S}ystems ({CHI 2016})}\
  (\BPGS\ 2371--3282).
\newblock
\APACaddressPublisher{San Jose}{ACM}.
\PrintBackRefs{\CurrentBib}

\bibitem [\protect \citeauthoryear {%
{European Commission}%
}{%
{European Commission}%
}{%
{\protect \APACyear {2020}}%
}]{%
EC20d}
\APACinsertmetastar {%
EC20d}%
\begin{APACrefauthors}%
{European Commission}%
\end{APACrefauthors}%
\unskip\
\newblock
\APACrefYearMonthDay{2020}{}{}.
\newblock
\APACrefbtitle {{D}igital {S}ervices {A}ct -- {D}eepening the {I}nternal
  {M}arket and {C}larifying {R}esponsibilities for {D}igital {S}ervices.}
  {{D}igital {S}ervices {A}ct -- {D}eepening the {I}nternal {M}arket and
  {C}larifying {R}esponsibilities for {D}igital {S}ervices.}
\newblock
\APACrefnote{{C}ontributions to the {O}pen {P}ublic {C}onsultation. Available
  online in February 2023:
  \url{https://ec.europa.eu/info/law/better-regulation/have-your-say/initiatives/12417-Digital-Services-Act-deepening-the-Internal-Market-and-clarifying-responsibilities-for-digital-services/public-consultation_en}}
\PrintBackRefs{\CurrentBib}

\bibitem [\protect \citeauthoryear {%
{European Commission}%
}{%
{European Commission}%
}{%
{\protect \APACyear {2023}}%
{\protect \APACexlab {{\protect \BCnt {1}}}}}]{%
EC23a}
\APACinsertmetastar {%
EC23a}%
\begin{APACrefauthors}%
{European Commission}%
\end{APACrefauthors}%
\unskip\
\newblock
\APACrefYearMonthDay{2023{\protect \BCnt {1}}}{}{}.
\newblock
\APACrefbtitle {{D}igital {S}ervices {A}ct: {A}pplication of the {R}isk
  {M}anagement {F}ramework to {R}ussian {D}isinformation {C}ampaigns.}
  {{D}igital {S}ervices {A}ct: {A}pplication of the {R}isk {M}anagement
  {F}ramework to {R}ussian {D}isinformation {C}ampaigns.}
\newblock
\APACrefnote{{D}irectorate-{G}eneral for {C}ommunications {N}etworks, {C}ontent
  and {T}echnology, {C}ontract {N}umber {PN}/2022/022. {A}vailable online in
  October:
  \url{https://op.europa.eu/en/publication-detail/-/publication/c1d645d0-42f5-11ee-a8b8-01aa75ed71a1/language-de}}
\PrintBackRefs{\CurrentBib}

\bibitem [\protect \citeauthoryear {%
{European Commission}%
}{%
{European Commission}%
}{%
{\protect \APACyear {2023}}%
{\protect \APACexlab {{\protect \BCnt {2}}}}}]{%
EC23b}
\APACinsertmetastar {%
EC23b}%
\begin{APACrefauthors}%
{European Commission}%
\end{APACrefauthors}%
\unskip\
\newblock
\APACrefYearMonthDay{2023{\protect \BCnt {2}}}{}{}.
\newblock
\APACrefbtitle {{EU} {I}nternet {F}orum: {S}tudy on the {R}ole and {E}ffects of
  the {U}se of {A}lgorithmic {A}mplification to {S}pread {T}errorist, {V}iolent
  {E}xtremist and {B}orderline {C}ontent.} {{EU} {I}nternet {F}orum: {S}tudy on
  the {R}ole and {E}ffects of the {U}se of {A}lgorithmic {A}mplification to
  {S}pread {T}errorist, {V}iolent {E}xtremist and {B}orderline {C}ontent.}
\newblock
\APACrefnote{{S}tudy {C}ommissioned by {DG HOME}, {D}irectorate-{G}eneral for
  {M}igration and {H}ome {A}ffairs. {A}vailable online in October:
  \url{https://op.europa.eu/en/publication-detail/-/publication/6818e404-7217-11ee-9220-01aa75ed71a1/language-en}}
\PrintBackRefs{\CurrentBib}

\bibitem [\protect \citeauthoryear {%
Flick%
}{%
Flick%
}{%
{\protect \APACyear {2004}}%
}]{%
Flick04}
\APACinsertmetastar {%
Flick04}%
\begin{APACrefauthors}%
Flick, U.%
\end{APACrefauthors}%
\unskip\
\newblock
\APACrefYearMonthDay{2004}{}{}.
\newblock
{\BBOQ}\APACrefatitle {{T}riangulation in {Q}ualitative {R}esearch}
  {{T}riangulation in {Q}ualitative {R}esearch}.{\BBCQ}
\newblock
 U.~Flick, E.~{von Kardoff}\BCBL {}\ \BBA {} I.~Steinke\ (\BEDS),
  \APACrefbtitle {{A} {C}ompanion to {Q}ualitative {R}esearch} {{A} {C}ompanion
  to {Q}ualitative {R}esearch}\ (\BPGS\ 178--183).
\newblock
\APACaddressPublisher{London}{SAGE}.
\PrintBackRefs{\CurrentBib}

\bibitem [\protect \citeauthoryear {%
Gedikli%
, Jannach%
\BCBL {}\ \BBA {} Ge%
}{%
Gedikli%
\ \protect \BOthers {.}}{%
{\protect \APACyear {2014}}%
}]{%
Gedikli14}
\APACinsertmetastar {%
Gedikli14}%
\begin{APACrefauthors}%
Gedikli, F.%
, Jannach, D.%
\BCBL {} Ge, M.%
\end{APACrefauthors}%
\unskip\
\newblock
\APACrefYearMonthDay{2014}{}{}.
\newblock
{\BBOQ}\APACrefatitle {{H}ow {S}hould {I} {E}xplain? {A} {C}omparison of
  {D}ifferent {E}xplanation {T}ypes for {R}ecommender {S}ystems} {{H}ow
  {S}hould {I} {E}xplain? {A} {C}omparison of {D}ifferent {E}xplanation {T}ypes
  for {R}ecommender {S}ystems}.{\BBCQ}
\newblock
\APACjournalVolNumPages{International Journal of Human-Computer
  Studies}{72}{4}{367--382}.
\newblock

\newblock

\PrintBackRefs{\CurrentBib}

\bibitem [\protect \citeauthoryear {%
Geissler%
, Maarouf%
\BCBL {}\ \BBA {} Feuerriegel%
}{%
Geissler%
\ \protect \BOthers {.}}{%
{\protect \APACyear {2023}}%
}]{%
Geissler23}
\APACinsertmetastar {%
Geissler23}%
\begin{APACrefauthors}%
Geissler, D.%
, Maarouf, A.%
\BCBL {} Feuerriegel, S.%
\end{APACrefauthors}%
\unskip\
\newblock
\APACrefYearMonthDay{2023}{}{}.
\newblock
\APACrefbtitle {{C}ausal {U}nderstanding of {W}hy {U}sers {S}hare {H}ate
  {S}peech on {S}ocial {M}edia.} {{C}ausal {U}nderstanding of {W}hy {U}sers
  {S}hare {H}ate {S}peech on {S}ocial {M}edia.}
\newblock
\APACrefnote{{A}rchived manuscript. {A}vailable online in October:
  \url{https://arxiv.org/abs/2310.15772}}
\PrintBackRefs{\CurrentBib}

\bibitem [\protect \citeauthoryear {%
Ghayda%
\ \protect \BOthers {.}}{%
Ghayda%
\ \protect \BOthers {.}}{%
{\protect \APACyear {2018}}%
}]{%
Ghayda18}
\APACinsertmetastar {%
Ghayda18}%
\begin{APACrefauthors}%
Ghayda, H.%
, an Alava~S\'eraphinc, B.S.%
, Divinad, F.%
, Lavoie, L.%
, Arbere, F.%
, Wynnpaulf, V.%
\BDBL {}Stijnh, S.%
\end{APACrefauthors}%
\unskip\
\newblock
\APACrefYearMonthDay{2018}{}{}.
\newblock
{\BBOQ}\APACrefatitle {{E}xposure to {E}xtremist {O}nline {C}ontent {C}ould
  {L}ead to {V}iolent {R}adicalization: {A} {S}ystematic {R}eview of
  {E}mpirical {E}vidence} {{E}xposure to {E}xtremist {O}nline {C}ontent {C}ould
  {L}ead to {V}iolent {R}adicalization: {A} {S}ystematic {R}eview of
  {E}mpirical {E}vidence}.{\BBCQ}
\newblock
\APACjournalVolNumPages{International Journal of Developmental
  Science}{12}{1--2}{71--88}.
\newblock

\newblock

\PrintBackRefs{\CurrentBib}

\bibitem [\protect \citeauthoryear {%
Gowder%
}{%
Gowder%
}{%
{\protect \APACyear {2023}}%
}]{%
Gowder23}
\APACinsertmetastar {%
Gowder23}%
\begin{APACrefauthors}%
Gowder, P.%
\end{APACrefauthors}%
\unskip\
\newblock
\APACrefYear{2023}.
\newblock
\APACrefbtitle {{T}he {N}etworked {L}eviathan for {D}emocratic {P}latforms}
  {{T}he {N}etworked {L}eviathan for {D}emocratic {P}latforms}.
\newblock
\APACaddressPublisher{Cambridge}{Cambridge University Press}.
\PrintBackRefs{\CurrentBib}

\bibitem [\protect \citeauthoryear {%
Guess%
\ \protect \BOthers {.}}{%
Guess%
\ \protect \BOthers {.}}{%
{\protect \APACyear {2023}}%
}]{%
Guess23}
\APACinsertmetastar {%
Guess23}%
\begin{APACrefauthors}%
Guess, A.M.%
, Malhotra, N.%
, Pan, J.%
, Barberá, P.%
, Allcott, H.%
, Brown, T.%
\BDBL {}Tucker, J.A.%
\end{APACrefauthors}%
\unskip\
\newblock
\APACrefYearMonthDay{2023}{}{}.
\newblock
{\BBOQ}\APACrefatitle {{H}ow {D}o {S}ocial {M}edia {F}eed {A}lgorithms {A}ffect
  {A}ttitudes and {B}ehavior in an {E}lection {C}ampaign?} {{H}ow {D}o {S}ocial
  {M}edia {F}eed {A}lgorithms {A}ffect {A}ttitudes and {B}ehavior in an
  {E}lection {C}ampaign?}{\BBCQ}
\newblock
\APACjournalVolNumPages{Science}{381}{6656}{398--404}.
\newblock

\newblock

\PrintBackRefs{\CurrentBib}

\bibitem [\protect \citeauthoryear {%
Heath%
\ \BBA {} Cowley%
}{%
Heath%
\ \BBA {} Cowley%
}{%
{\protect \APACyear {2004}}%
}]{%
Heath04}
\APACinsertmetastar {%
Heath04}%
\begin{APACrefauthors}%
Heath, H.%
\BCBT {}\ \BBA {} Cowley, S.%
\end{APACrefauthors}%
\unskip\
\newblock
\APACrefYearMonthDay{2004}{}{}.
\newblock
{\BBOQ}\APACrefatitle {{D}eveloping a {G}rounded {T}heory {A}pproach: {A}
  {C}omparison of {G}laser and {S}traxbuss} {{D}eveloping a {G}rounded {T}heory
  {A}pproach: {A} {C}omparison of {G}laser and {S}traxbuss}.{\BBCQ}
\newblock
\APACjournalVolNumPages{Nursing Studies}{41}{}{141--150}.
\newblock

\newblock

\PrintBackRefs{\CurrentBib}

\bibitem [\protect \citeauthoryear {%
Helberger%
, {van Drunen}%
, Vrijenhoek%
\BCBL {}\ \BBA {} M\"oller%
}{%
Helberger%
\ \protect \BOthers {.}}{%
{\protect \APACyear {2021}}%
}]{%
Helberger21}
\APACinsertmetastar {%
Helberger21}%
\begin{APACrefauthors}%
Helberger, N.%
, {van Drunen}, M.%
, Vrijenhoek, S.%
\BCBL {} M\"oller, J.%
\end{APACrefauthors}%
\unskip\
\newblock
\APACrefYearMonthDay{2021}{}{}.
\newblock
{\BBOQ}\APACrefatitle {{R}egulation of {N}ews {R}ecommenders in the {D}igital
  {S}ervices {A}ct: {E}mpowering {D}avid {A}gainst the {V}ery {L}arge {O}nline
  {G}oliath} {{R}egulation of {N}ews {R}ecommenders in the {D}igital {S}ervices
  {A}ct: {E}mpowering {D}avid {A}gainst the {V}ery {L}arge {O}nline
  {G}oliath}.{\BBCQ}
\newblock
\APACjournalVolNumPages{Internet Policy Review}{}{}{}.
\newblock
\APACrefnote{{O}pinion. {A}vailable online in February 2023:
  \url{https://policyreview.info/articles/news/regulation-news-recommenders-digital-services-act-empowering-david-against-very-large}}
\newblock

\newblock

\PrintBackRefs{\CurrentBib}

\bibitem [\protect \citeauthoryear {%
Heldt%
}{%
Heldt%
}{%
{\protect \APACyear {2022}}%
}]{%
Heldt22}
\APACinsertmetastar {%
Heldt22}%
\begin{APACrefauthors}%
Heldt, A.P.%
\end{APACrefauthors}%
\unskip\
\newblock
\APACrefYearMonthDay{2022}{}{}.
\newblock
{\BBOQ}\APACrefatitle {{EU} {D}igital {S}ervices {A}ct: {T}he {W}hite {H}ope of
  {I}ntermediary {R}egulation} {{EU} {D}igital {S}ervices {A}ct: {T}he {W}hite
  {H}ope of {I}ntermediary {R}egulation}.{\BBCQ}
\newblock
 T.~Flew\ \BBA {} F.R.~Martin\ (\BEDS), \APACrefbtitle {{D}igital {P}latform
  {R}egulation: {G}lobal {P}erspectives on {I}nternet {G}overnance} {{D}igital
  {P}latform {R}egulation: {G}lobal {P}erspectives on {I}nternet {G}overnance}\
  (\BPGS\ 69--84).
\newblock
\APACaddressPublisher{Cham}{Palgrave Macmillan}.
\PrintBackRefs{\CurrentBib}

\bibitem [\protect \citeauthoryear {%
Hildebrandt%
}{%
Hildebrandt%
}{%
{\protect \APACyear {2009}}%
}]{%
Hildebrandt09}
\APACinsertmetastar {%
Hildebrandt09}%
\begin{APACrefauthors}%
Hildebrandt, M.%
\end{APACrefauthors}%
\unskip\
\newblock
\APACrefYearMonthDay{2009}{}{}.
\newblock
{\BBOQ}\APACrefatitle {{W}ho is {P}rofiling {W}ho? {I}nvisible {V}isibility}
  {{W}ho is {P}rofiling {W}ho? {I}nvisible {V}isibility}.{\BBCQ}
\newblock
 S.~Gutwirth, Y.~Poullet, P.D.~Hert, C.~{de Terwangne}\BCBL {}\ \BBA {}
  S.~Nouwt\ (\BEDS), \APACrefbtitle {{R}einventing {D}ata {P}rotection?}
  {{R}einventing {D}ata {P}rotection?}\ (\BPGS\ 239--252).
\newblock
\APACaddressPublisher{Cham}{Springer}.
\PrintBackRefs{\CurrentBib}

\bibitem [\protect \citeauthoryear {%
Hildebrandt%
}{%
Hildebrandt%
}{%
{\protect \APACyear {2020}}%
}]{%
Hildebrandt20}
\APACinsertmetastar {%
Hildebrandt20}%
\begin{APACrefauthors}%
Hildebrandt, M.%
\end{APACrefauthors}%
\unskip\
\newblock
\APACrefYear{2020}.
\newblock
\APACrefbtitle {{L}aw for {C}omputer {S}cientists and {O}ther {F}olk} {{L}aw
  for {C}omputer {S}cientists and {O}ther {F}olk}.
\newblock
\APACaddressPublisher{Oxford}{Oxford University Press}.
\PrintBackRefs{\CurrentBib}

\bibitem [\protect \citeauthoryear {%
Hjerppe%
, Ruohonen%
\BCBL {}\ \BBA {} Lepp\"anen%
}{%
Hjerppe%
\ \protect \BOthers {.}}{%
{\protect \APACyear {2023}}%
}]{%
Hjerppe23SOSYM}
\APACinsertmetastar {%
Hjerppe23SOSYM}%
\begin{APACrefauthors}%
Hjerppe, K.%
, Ruohonen, J.%
\BCBL {} Lepp\"anen, V.%
\end{APACrefauthors}%
\unskip\
\newblock
\APACrefYearMonthDay{2023}{}{}.
\newblock
{\BBOQ}\APACrefatitle {{E}xtracting {LPL} {P}rivacy {P}olicy {P}urposes from
  {A}nnotated {W}eb {S}ervice {S}ource {C}ode} {{E}xtracting {LPL} {P}rivacy
  {P}olicy {P}urposes from {A}nnotated {W}eb {S}ervice {S}ource {C}ode}.{\BBCQ}
\newblock
\APACjournalVolNumPages{Software and Systems Modeling}{22}{}{331--349}.
\newblock

\newblock

\PrintBackRefs{\CurrentBib}

\bibitem [\protect \citeauthoryear {%
Hsieh%
\ \BBA {} Shannon%
}{%
Hsieh%
\ \BBA {} Shannon%
}{%
{\protect \APACyear {2005}}%
}]{%
Hsieh05}
\APACinsertmetastar {%
Hsieh05}%
\begin{APACrefauthors}%
Hsieh, H\BHBI F.%
\BCBT {}\ \BBA {} Shannon, S.E.%
\end{APACrefauthors}%
\unskip\
\newblock
\APACrefYearMonthDay{2005}{}{}.
\newblock
{\BBOQ}\APACrefatitle {{T}hree {A}pproaches to {Q}ualitative {C}ontent
  {A}nalysis} {{T}hree {A}pproaches to {Q}ualitative {C}ontent
  {A}nalysis}.{\BBCQ}
\newblock
\APACjournalVolNumPages{Qualitative Health Research}{15}{9}{1277--1288}.
\newblock

\newblock

\PrintBackRefs{\CurrentBib}

\bibitem [\protect \citeauthoryear {%
Humberstone%
}{%
Humberstone%
}{%
{\protect \APACyear {2023}}%
}]{%
Humberstone23}
\APACinsertmetastar {%
Humberstone23}%
\begin{APACrefauthors}%
Humberstone, T.%
\end{APACrefauthors}%
\unskip\
\newblock
\APACrefYearMonthDay{2023}{}{}.
\newblock
\APACrefbtitle {{I}’m a {L}uddite (and {S}o {C}an {Y}ou!).} {{I}’m a
  {L}uddite (and {S}o {C}an {Y}ou!).}
\newblock
\APACrefnote{{T}he {N}ib. {A}vailable online in March 2024:
  \url{https://thenib.com/im-a-luddite/}}
\PrintBackRefs{\CurrentBib}

\bibitem [\protect \citeauthoryear {%
Iandoli%
, Primario%
\BCBL {}\ \BBA {} Zollo%
}{%
Iandoli%
\ \protect \BOthers {.}}{%
{\protect \APACyear {2021}}%
}]{%
Iandoli21}
\APACinsertmetastar {%
Iandoli21}%
\begin{APACrefauthors}%
Iandoli, L.%
, Primario, S.%
\BCBL {} Zollo, G.%
\end{APACrefauthors}%
\unskip\
\newblock
\APACrefYearMonthDay{2021}{}{}.
\newblock
{\BBOQ}\APACrefatitle {{T}he {I}mpact of {G}roup {P}olarization on the
  {Q}uality of {O}nline {D}ebate in {S}ocial {M}edia: {A} {S}ystematic
  {L}iterature {R}eview} {{T}he {I}mpact of {G}roup {P}olarization on the
  {Q}uality of {O}nline {D}ebate in {S}ocial {M}edia: {A} {S}ystematic
  {L}iterature {R}eview}.{\BBCQ}
\newblock
\APACjournalVolNumPages{Technological Forecasting and Social
  Change}{170}{}{120924}.
\newblock

\newblock

\PrintBackRefs{\CurrentBib}

\bibitem [\protect \citeauthoryear {%
{ICCL}%
}{%
{ICCL}%
}{%
{\protect \APACyear {2023}}%
}]{%
ICCL23}
\APACinsertmetastar {%
ICCL23}%
\begin{APACrefauthors}%
{ICCL}%
\end{APACrefauthors}%
\unskip\
\newblock
\APACrefYearMonthDay{2023}{}{}.
\newblock
\APACrefbtitle {{T}he {E}uropean {C}ommission {M}ust {F}ollow {I}reland’s
  {L}ead, and {S}witch off {B}ig {T}ech’s {T}oxic {A}lgorithms.} {{T}he
  {E}uropean {C}ommission {M}ust {F}ollow {I}reland’s {L}ead, and {S}witch
  off {B}ig {T}ech’s {T}oxic {A}lgorithms.}
\newblock
\APACrefnote{{I}rish {C}ouncil for {C}ivil {L}iberties. {A}vailable online in
  December:
  \url{https://www.iccl.ie/2023/the-european-commission-must-follow-irelands-lead-and-switch-off-big-techs-toxic-algorithms/}}
\PrintBackRefs{\CurrentBib}

\bibitem [\protect \citeauthoryear {%
Iqbal%
\ \protect \BOthers {.}}{%
Iqbal%
\ \protect \BOthers {.}}{%
{\protect \APACyear {2023}}%
}]{%
Iqbal23}
\APACinsertmetastar {%
Iqbal23}%
\begin{APACrefauthors}%
Iqbal, U.%
, Bahrami, P.N.%
, Trimananda, R.%
, Cui, H.%
, {Gamero-Garrido}, A.%
, Dubois, D.%
\BDBL {}Shafiq, Z.%
\end{APACrefauthors}%
\unskip\
\newblock
\APACrefYearMonthDay{2023}{}{}.
\newblock
{\BBOQ}\APACrefatitle {{T}racking, {P}rofiling, and {A}d {T}argeting in the
  {A}lexa {E}cho {S}mart {S}peaker {E}cosystem} {{T}racking, {P}rofiling, and
  {A}d {T}argeting in the {A}lexa {E}cho {S}mart {S}peaker {E}cosystem}.{\BBCQ}
\newblock
 \APACrefbtitle {{P}roceedings of the {I}nternet {M}easurement {C}onference
  ({IMC} 2023).} {{P}roceedings of the {I}nternet {M}easurement {C}onference
  ({IMC} 2023).}
\newblock
\APACaddressPublisher{Montreal}{ACM}.
\PrintBackRefs{\CurrentBib}

\bibitem [\protect \citeauthoryear {%
Jackson%
\ \BBA {} Malaret%
}{%
Jackson%
\ \BBA {} Malaret%
}{%
{\protect \APACyear {2023}}%
}]{%
DFRLab23a}
\APACinsertmetastar {%
DFRLab23a}%
\begin{APACrefauthors}%
Jackson, R.%
\BCBT {}\ \BBA {} Malaret, J.%
\end{APACrefauthors}%
\unskip\
\newblock
\APACrefYearMonthDay{2023}{}{}.
\newblock
\APACrefbtitle {{A}s {I}srael and {H}amas {G}o to {W}ar, the {D}igital
  {S}ervices {A}ct {F}aces {I}ts {F}irst {M}ajor {T}est.} {{A}s {I}srael and
  {H}amas {G}o to {W}ar, the {D}igital {S}ervices {A}ct {F}aces {I}ts {F}irst
  {M}ajor {T}est.}
\newblock
\APACrefnote{{T}he {D}igital {F}orensic {R}esearch {L}ab {(DFRLab)}, the
  {A}tlantic {C}ouncil. Available online in October:
  \url{https://dfrlab.org/2023/10/26/as-israel-and-hamas-go-to-war-the-digital-services-act-faces-its-first-major-test/}}
\PrintBackRefs{\CurrentBib}

\bibitem [\protect \citeauthoryear {%
Kalmus%
, Bolin%
\BCBL {}\ \BBA {} Figueiras%
}{%
Kalmus%
\ \protect \BOthers {.}}{%
{\protect \APACyear {2022}}%
}]{%
Kalmus22}
\APACinsertmetastar {%
Kalmus22}%
\begin{APACrefauthors}%
Kalmus, V.%
, Bolin, G.%
\BCBL {} Figueiras, R.%
\end{APACrefauthors}%
\unskip\
\newblock
\APACrefYearMonthDay{2022}{}{}.
\newblock
{\BBOQ}\APACrefatitle {{W}ho {I}s {A}fraid of {D}ataveillance? {A}ttitudes
  {T}oward {O}nline {S}urveillance in a {C}ross-{C}ultural and {G}enerational
  {P}erspective} {{W}ho {I}s {A}fraid of {D}ataveillance? {A}ttitudes {T}oward
  {O}nline {S}urveillance in a {C}ross-{C}ultural and {G}enerational
  {P}erspective}.{\BBCQ}
\newblock
\APACjournalVolNumPages{New Media \& Society}{}{Published online in
  November}{1--23}.
\newblock

\newblock

\PrintBackRefs{\CurrentBib}

\bibitem [\protect \citeauthoryear {%
Kapoor%
\ \BBA {} Narayanan%
}{%
Kapoor%
\ \BBA {} Narayanan%
}{%
{\protect \APACyear {2023}}%
}]{%
Kapoor23}
\APACinsertmetastar {%
Kapoor23}%
\begin{APACrefauthors}%
Kapoor, S.%
\BCBT {}\ \BBA {} Narayanan, A.%
\end{APACrefauthors}%
\unskip\
\newblock
\APACrefYearMonthDay{2023}{}{}.
\newblock
\APACrefbtitle {{H}ow to {P}repare for the {D}eluge of {G}enerative {AI} on
  {S}ocial {M}edia: {A} {G}rounded {A}nalysis of the {C}hallenges and
  {O}pportunities.} {{H}ow to {P}repare for the {D}eluge of {G}enerative {AI}
  on {S}ocial {M}edia: {A} {G}rounded {A}nalysis of the {C}hallenges and
  {O}pportunities.}
\newblock
\APACrefnote{{K}night {F}irst {A}mendment {I}nstitute at {C}olumbia
  {U}niversity. {A}vailable online in October:
  \url{https://knightcolumbia.org/content/how-to-prepare-for-the-deluge-of-generative-ai-on-social-media}}
\PrintBackRefs{\CurrentBib}

\bibitem [\protect \citeauthoryear {%
Kieslich%
, Keller%
\BCBL {}\ \BBA {} Starke%
}{%
Kieslich%
\ \protect \BOthers {.}}{%
{\protect \APACyear {2022}}%
}]{%
Kieslich22}
\APACinsertmetastar {%
Kieslich22}%
\begin{APACrefauthors}%
Kieslich, K.%
, Keller, B.%
\BCBL {} Starke, C.%
\end{APACrefauthors}%
\unskip\
\newblock
\APACrefYearMonthDay{2022}{}{}.
\newblock
{\BBOQ}\APACrefatitle {{A}rtificial {I}ntelligence {E}thics by {D}esign.
  {E}valuating {P}ublic {P}erception on the {I}mportance of {E}thical {D}esign
  {P}rinciples of {A}rtificial {I}ntelligence} {{A}rtificial {I}ntelligence
  {E}thics by {D}esign. {E}valuating {P}ublic {P}erception on the {I}mportance
  of {E}thical {D}esign {P}rinciples of {A}rtificial {I}ntelligence}.{\BBCQ}
\newblock
\APACjournalVolNumPages{Big Data \& Society}{9}{1}{1--15}.
\newblock

\newblock

\PrintBackRefs{\CurrentBib}

\bibitem [\protect \citeauthoryear {%
Kieslich%
, L\"unich%
\BCBL {}\ \BBA {} Marcinkowski%
}{%
Kieslich%
\ \protect \BOthers {.}}{%
{\protect \APACyear {2021}}%
}]{%
Kieslich21}
\APACinsertmetastar {%
Kieslich21}%
\begin{APACrefauthors}%
Kieslich, K.%
, L\"unich, M.%
\BCBL {} Marcinkowski, F.%
\end{APACrefauthors}%
\unskip\
\newblock
\APACrefYearMonthDay{2021}{}{}.
\newblock
{\BBOQ}\APACrefatitle {{T}he {T}hreats of {A}rtificial {I}ntelligence {S}cale
  {(TAI)}. {D}evelopment, {M}easurement and {T}est {O}ver {T}hree {A}pplication
  {D}omains} {{T}he {T}hreats of {A}rtificial {I}ntelligence {S}cale {(TAI)}.
  {D}evelopment, {M}easurement and {T}est {O}ver {T}hree {A}pplication
  {D}omains}.{\BBCQ}
\newblock
\APACjournalVolNumPages{International Journal of Social
  Robotics}{13}{}{1563--1577}.
\newblock

\newblock

\PrintBackRefs{\CurrentBib}

\bibitem [\protect \citeauthoryear {%
Knijnenburg%
, Willemsen%
\BCBL {}\ \BBA {} Kobsa%
}{%
Knijnenburg%
\ \protect \BOthers {.}}{%
{\protect \APACyear {2011}}%
}]{%
Knijnenburg11}
\APACinsertmetastar {%
Knijnenburg11}%
\begin{APACrefauthors}%
Knijnenburg, B.P.%
, Willemsen, M.C.%
\BCBL {} Kobsa, A.%
\end{APACrefauthors}%
\unskip\
\newblock
\APACrefYearMonthDay{2011}{}{}.
\newblock
{\BBOQ}\APACrefatitle {{A} {P}ragmatic {P}rocedure to {S}upport the
  {U}ser-{C}entric {E}valuation of {R}ecommender {S}ystems} {{A} {P}ragmatic
  {P}rocedure to {S}upport the {U}ser-{C}entric {E}valuation of {R}ecommender
  {S}ystems}.{\BBCQ}
\newblock
 \APACrefbtitle {{P}roceedings of the 11th {ACM} {I}nternational {C}onference
  on {R}ecommender {S}ystems ({R}ec{S}ys 2011)} {{P}roceedings of the 11th
  {ACM} {I}nternational {C}onference on {R}ecommender {S}ystems ({R}ec{S}ys
  2011)}\ (\BPGS\ 321--324).
\newblock
\APACaddressPublisher{Chicago}{ACM}.
\PrintBackRefs{\CurrentBib}

\bibitem [\protect \citeauthoryear {%
Kubin%
\ \BBA {} {von Sikorski}%
}{%
Kubin%
\ \BBA {} {von Sikorski}%
}{%
{\protect \APACyear {2021}}%
}]{%
Kubin21}
\APACinsertmetastar {%
Kubin21}%
\begin{APACrefauthors}%
Kubin, E.%
\BCBT {}\ \BBA {} {von Sikorski}, C.%
\end{APACrefauthors}%
\unskip\
\newblock
\APACrefYearMonthDay{2021}{}{}.
\newblock
{\BBOQ}\APACrefatitle {{T}he {R}ole of ({S}ocial) {M}edia in {P}olitical
  {P}olarization: {A} {S}ystematic {R}eview} {{T}he {R}ole of ({S}ocial)
  {M}edia in {P}olitical {P}olarization: {A} {S}ystematic {R}eview}.{\BBCQ}
\newblock
\APACjournalVolNumPages{Annals of the International Communication
  Association}{45}{3}{188--206}.
\newblock

\newblock

\PrintBackRefs{\CurrentBib}

\bibitem [\protect \citeauthoryear {%
Kupiec%
}{%
Kupiec%
}{%
{\protect \APACyear {2023}}%
}]{%
Kupiec23}
\APACinsertmetastar {%
Kupiec23}%
\begin{APACrefauthors}%
Kupiec, M.%
\end{APACrefauthors}%
\unskip\
\newblock
\APACrefYearMonthDay{2023}{}{}.
\newblock
\APACrefbtitle {{A}mazon’s {I}nterim {R}elief to {S}uspend {O}bligations on
  {O}nline {A}dvertising {T}ransparency {U}nder the {DSA}: {O}ne {S}wallow
  {D}oesn’t {M}ake a {S}ummer.} {{A}mazon’s {I}nterim {R}elief to {S}uspend
  {O}bligations on {O}nline {A}dvertising {T}ransparency {U}nder the {DSA}:
  {O}ne {S}wallow {D}oesn’t {M}ake a {S}ummer.}
\newblock
\APACrefnote{{K}luwer {C}ompetition {L}aw {B}log. {A}vailable online in
  December:
  \url{https://competitionlawblog.kluwercompetitionlaw.com/2023/11/06/amazons-interim-relief-to-suspend-obligations-on-online-advertising-transparency-under-the-dsa-one-swallow-doesnt-make-a-summer/}}
\PrintBackRefs{\CurrentBib}

\bibitem [\protect \citeauthoryear {%
Lai%
, Chen%
, Liao%
, {Smith-Renner}%
\BCBL {}\ \BBA {} Tan%
}{%
Lai%
\ \protect \BOthers {.}}{%
{\protect \APACyear {2021}}%
}]{%
Lai21}
\APACinsertmetastar {%
Lai21}%
\begin{APACrefauthors}%
Lai, V.%
, Chen, C.%
, Liao, Q.V.%
, {Smith-Renner}, A.%
\BCBL {} Tan, C.%
\end{APACrefauthors}%
\unskip\
\newblock
\APACrefYearMonthDay{2021}{}{}.
\newblock
\APACrefbtitle {{T}owards a {S}cience of {H}uman-{AI} {D}ecision {M}aking: {A}
  {S}urvey of {E}mpirical {S}tudies.} {{T}owards a {S}cience of {H}uman-{AI}
  {D}ecision {M}aking: {A} {S}urvey of {E}mpirical {S}tudies.}
\newblock
\APACrefnote{{A}rchived manuscript. {A}vailable online in February 2023:
  \url{https://arxiv.org/abs/2112.11471}}
\PrintBackRefs{\CurrentBib}

\bibitem [\protect \citeauthoryear {%
Laperdrix%
, Bielova%
\BCBL {}\ \BBA {} Avoine%
}{%
Laperdrix%
\ \protect \BOthers {.}}{%
{\protect \APACyear {2020}}%
}]{%
Laperdrix20}
\APACinsertmetastar {%
Laperdrix20}%
\begin{APACrefauthors}%
Laperdrix, P.%
, Bielova, N.%
\BCBL {} Avoine, B.B.G.%
\end{APACrefauthors}%
\unskip\
\newblock
\APACrefYearMonthDay{2020}{}{}.
\newblock
{\BBOQ}\APACrefatitle {{B}rowser {F}ingerprinting: {A} {S}urvey} {{B}rowser
  {F}ingerprinting: {A} {S}urvey}.{\BBCQ}
\newblock
\APACjournalVolNumPages{ACM Transactions on the Web}{14}{2}{1--33}.
\newblock

\newblock

\PrintBackRefs{\CurrentBib}

\bibitem [\protect \citeauthoryear {%
Laufer%
\ \BBA {} Nissenbaum%
}{%
Laufer%
\ \BBA {} Nissenbaum%
}{%
{\protect \APACyear {2023}}%
}]{%
Laufer23}
\APACinsertmetastar {%
Laufer23}%
\begin{APACrefauthors}%
Laufer, B.%
\BCBT {}\ \BBA {} Nissenbaum, H.%
\end{APACrefauthors}%
\unskip\
\newblock
\APACrefYearMonthDay{2023}{}{}.
\newblock
\APACrefbtitle {{A}lgorithmic {D}isplacement of {S}ocial {T}rust.}
  {{A}lgorithmic {D}isplacement of {S}ocial {T}rust.}
\newblock
\APACrefnote{{K}night {F}irst {A}mendment {I}nstitute at {C}olumbia
  {U}niversity. {A}vailable online in December:
  \url{https://knightcolumbia.org/content/algorithmic-displacement-of-social-trust}}
\PrintBackRefs{\CurrentBib}

\bibitem [\protect \citeauthoryear {%
Lee%
}{%
Lee%
}{%
{\protect \APACyear {2018}}%
}]{%
Lee18}
\APACinsertmetastar {%
Lee18}%
\begin{APACrefauthors}%
Lee, M.K.%
\end{APACrefauthors}%
\unskip\
\newblock
\APACrefYearMonthDay{2018}{}{}.
\newblock
{\BBOQ}\APACrefatitle {{U}nderstanding {P}erception of {A}lgorithmic
  {D}ecisions: {F}airness, {T}rust, and {E}motion in {R}esponse to
  {A}lgorithmic {M}anagement} {{U}nderstanding {P}erception of {A}lgorithmic
  {D}ecisions: {F}airness, {T}rust, and {E}motion in {R}esponse to
  {A}lgorithmic {M}anagement}.{\BBCQ}
\newblock
\APACjournalVolNumPages{Big Data \& Society}{5}{1}{2053951718756684}.
\newblock

\newblock

\PrintBackRefs{\CurrentBib}

\bibitem [\protect \citeauthoryear {%
Liu%
, Iqbal%
\BCBL {}\ \BBA {} Saxena%
}{%
Liu%
\ \protect \BOthers {.}}{%
{\protect \APACyear {2022}}%
}]{%
LiuIqbal22}
\APACinsertmetastar {%
LiuIqbal22}%
\begin{APACrefauthors}%
Liu, Z.%
, Iqbal, U.%
\BCBL {} Saxena, N.%
\end{APACrefauthors}%
\unskip\
\newblock
\APACrefYearMonthDay{2022}{}{}.
\newblock
\APACrefbtitle {{O}pted {O}ut, {Y}et {T}racked: {A}re {R}egulations {E}nough to
  {P}rotect {Y}our {P}rivacy?} {{O}pted {O}ut, {Y}et {T}racked: {A}re
  {R}egulations {E}nough to {P}rotect {Y}our {P}rivacy?}
\newblock
\APACrefnote{{U}npublished manuscript. Available online in March 2023:
  \url{https://arxiv.org/abs/2202.00885}}
\PrintBackRefs{\CurrentBib}

\bibitem [\protect \citeauthoryear {%
Martijn%
, Conati%
\BCBL {}\ \BBA {} Verbert%
}{%
Martijn%
\ \protect \BOthers {.}}{%
{\protect \APACyear {2022}}%
}]{%
Martijn22}
\APACinsertmetastar {%
Martijn22}%
\begin{APACrefauthors}%
Martijn, M.%
, Conati, C.%
\BCBL {} Verbert, K.%
\end{APACrefauthors}%
\unskip\
\newblock
\APACrefYearMonthDay{2022}{}{}.
\newblock
{\BBOQ}\APACrefatitle {``{K}nowing {M}e, {K}nowing {Y}ou'': {P}ersonalized
  {E}xplanations for a {M}usic {R}ecommender {S}ystem} {``{K}nowing {M}e,
  {K}nowing {Y}ou'': {P}ersonalized {E}xplanations for a {M}usic {R}ecommender
  {S}ystem}.{\BBCQ}
\newblock
\APACjournalVolNumPages{User Modeling and User-Adapted
  Interaction}{32}{}{215--252}.
\newblock

\newblock

\PrintBackRefs{\CurrentBib}

\bibitem [\protect \citeauthoryear {%
Mcsherry%
}{%
Mcsherry%
}{%
{\protect \APACyear {2005}}%
}]{%
McSherry05}
\APACinsertmetastar {%
McSherry05}%
\begin{APACrefauthors}%
Mcsherry, D.%
\end{APACrefauthors}%
\unskip\
\newblock
\APACrefYearMonthDay{2005}{}{}.
\newblock
{\BBOQ}\APACrefatitle {{E}xplanation in {R}ecommender {S}ystems} {{E}xplanation
  in {R}ecommender {S}ystems}.{\BBCQ}
\newblock
\APACjournalVolNumPages{Artificial Intelligence Review}{24}{}{179--197}.
\newblock

\newblock

\PrintBackRefs{\CurrentBib}

\bibitem [\protect \citeauthoryear {%
Milano%
, Taddeo%
\BCBL {}\ \BBA {} Floridi%
}{%
Milano%
\ \protect \BOthers {.}}{%
{\protect \APACyear {2020}}%
}]{%
Milano20}
\APACinsertmetastar {%
Milano20}%
\begin{APACrefauthors}%
Milano, S.%
, Taddeo, M.%
\BCBL {} Floridi, L.%
\end{APACrefauthors}%
\unskip\
\newblock
\APACrefYearMonthDay{2020}{}{}.
\newblock
{\BBOQ}\APACrefatitle {{R}ecommender {S}ystems and {T}heir {E}thical
  {C}hallenges} {{R}ecommender {S}ystems and {T}heir {E}thical
  {C}hallenges}.{\BBCQ}
\newblock
\APACjournalVolNumPages{AI \& SOCIETY}{35}{}{957--967}.
\newblock

\newblock

\PrintBackRefs{\CurrentBib}

\bibitem [\protect \citeauthoryear {%
Morgan%
}{%
Morgan%
}{%
{\protect \APACyear {1993}}%
}]{%
Morgan93}
\APACinsertmetastar {%
Morgan93}%
\begin{APACrefauthors}%
Morgan, D.L.%
\end{APACrefauthors}%
\unskip\
\newblock
\APACrefYearMonthDay{1993}{}{}.
\newblock
{\BBOQ}\APACrefatitle {{Q}ualitative {C}ontent {A}nalysis: {A} {G}uide to
  {P}aths not {T}aken} {{Q}ualitative {C}ontent {A}nalysis: {A} {G}uide to
  {P}aths not {T}aken}.{\BBCQ}
\newblock
\APACjournalVolNumPages{Qualitative Health Research}{3}{1}{112--121}.
\newblock

\newblock

\PrintBackRefs{\CurrentBib}

\bibitem [\protect \citeauthoryear {%
Pagano%
}{%
Pagano%
}{%
{\protect \APACyear {2018}}%
}]{%
Pagano18}
\APACinsertmetastar {%
Pagano18}%
\begin{APACrefauthors}%
Pagano, N.A.%
\end{APACrefauthors}%
\unskip\
\newblock
\APACrefYearMonthDay{2018}{}{}.
\newblock
{\BBOQ}\APACrefatitle {{T}he {I}ndecency of the {C}ommunications {D}ecency
  {A}ct § 230: {U}njust {I}mmunity for {M}onstrous {S}ocial {M}edia
  {P}latforms} {{T}he {I}ndecency of the {C}ommunications {D}ecency {A}ct §
  230: {U}njust {I}mmunity for {M}onstrous {S}ocial {M}edia
  {P}latforms}.{\BBCQ}
\newblock
\APACjournalVolNumPages{Pace Law Review}{39}{1}{511--538}.
\newblock

\newblock

\PrintBackRefs{\CurrentBib}

\bibitem [\protect \citeauthoryear {%
Papaevangelou%
}{%
Papaevangelou%
}{%
{\protect \APACyear {2023}}%
}]{%
Papaevangelou23}
\APACinsertmetastar {%
Papaevangelou23}%
\begin{APACrefauthors}%
Papaevangelou, C.%
\end{APACrefauthors}%
\unskip\
\newblock
\APACrefYearMonthDay{2023}{}{}.
\newblock
{\BBOQ}\APACrefatitle {'{T}he {N}on-{I}nterference {P}rinciple': {D}ebating
  {O}nline {P}latforms' {T}reatment of {E}ditorial {C}ontent in the {E}uropean
  {U}nion's {D}igital {S}ervices {A}ct} {'{T}he {N}on-{I}nterference
  {P}rinciple': {D}ebating {O}nline {P}latforms' {T}reatment of {E}ditorial
  {C}ontent in the {E}uropean {U}nion's {D}igital {S}ervices {A}ct}.{\BBCQ}
\newblock
\APACjournalVolNumPages{European Journal of Communication}{38}{5}{466--483}.
\newblock

\newblock

\PrintBackRefs{\CurrentBib}

\bibitem [\protect \citeauthoryear {%
Patel%
}{%
Patel%
}{%
{\protect \APACyear {2023}}%
}]{%
Patel23}
\APACinsertmetastar {%
Patel23}%
\begin{APACrefauthors}%
Patel, N.%
\end{APACrefauthors}%
\unskip\
\newblock
\APACrefYearMonthDay{2023}{}{}.
\newblock
\APACrefbtitle {{H}arvard {P}rofessor {L}awrence {L}essig on {W}hy {AI} and
  {S}ocial {M}edia are {C}ausing a {F}ree {S}peech {C}risis for the
  {I}nternet.} {{H}arvard {P}rofessor {L}awrence {L}essig on {W}hy {AI} and
  {S}ocial {M}edia are {C}ausing a {F}ree {S}peech {C}risis for the
  {I}nternet.}
\newblock
\APACrefnote{{T}he {V}erge. {A}vailable online in October:
  \url{https://www.theverge.com/23929233/lawrence-lessig-free-speech-first-amendment-ai-content-moderation-decoder-interview}}
\PrintBackRefs{\CurrentBib}

\bibitem [\protect \citeauthoryear {%
Pu%
, Chen%
\BCBL {}\ \BBA {} Hu%
}{%
Pu%
\ \protect \BOthers {.}}{%
{\protect \APACyear {2012}}%
}]{%
Pu12}
\APACinsertmetastar {%
Pu12}%
\begin{APACrefauthors}%
Pu, P.%
, Chen, L.%
\BCBL {} Hu, R.%
\end{APACrefauthors}%
\unskip\
\newblock
\APACrefYearMonthDay{2012}{}{}.
\newblock
{\BBOQ}\APACrefatitle {{E}valuating {R}ecommender {S}ystems from the {U}ser's
  {P}erspective: {S}urvey of the {S}tate of the {A}rt} {{E}valuating
  {R}ecommender {S}ystems from the {U}ser's {P}erspective: {S}urvey of the
  {S}tate of the {A}rt}.{\BBCQ}
\newblock
\APACjournalVolNumPages{User Modeling and User-Adapted
  Interaction}{22}{}{317--355}.
\newblock

\newblock

\PrintBackRefs{\CurrentBib}

\bibitem [\protect \citeauthoryear {%
Ribeiro%
, Ottoni%
, West%
, Almeida%
\BCBL {}\ \BBA {} Meira%
}{%
Ribeiro%
\ \protect \BOthers {.}}{%
{\protect \APACyear {2020}}%
}]{%
Ribeiro20}
\APACinsertmetastar {%
Ribeiro20}%
\begin{APACrefauthors}%
Ribeiro, M.H.%
, Ottoni, R.%
, West, R.%
, Almeida, V.A.F.%
\BCBL {} Meira, W.%
\end{APACrefauthors}%
\unskip\
\newblock
\APACrefYearMonthDay{2020}{}{}.
\newblock
{\BBOQ}\APACrefatitle {{A}uditing {R}adicalization {P}athways on {Y}ou{T}ube}
  {{A}uditing {R}adicalization {P}athways on {Y}ou{T}ube}.{\BBCQ}
\newblock
 \APACrefbtitle {{P}roceedings of the {C}onference on {F}airness,
  {A}ccountability, and {T}ransparency ({FAT}* 2020)} {{P}roceedings of the
  {C}onference on {F}airness, {A}ccountability, and {T}ransparency ({FAT}*
  2020)}\ (\BPGS\ 131--141).
\newblock
\APACaddressPublisher{Barcelona}{ACM}.
\PrintBackRefs{\CurrentBib}

\bibitem [\protect \citeauthoryear {%
Riedenstein%
\ \BBA {} Echikson%
}{%
Riedenstein%
\ \BBA {} Echikson%
}{%
{\protect \APACyear {2023}}%
}]{%
CEPA23a}
\APACinsertmetastar {%
CEPA23a}%
\begin{APACrefauthors}%
Riedenstein, C.%
\BCBT {}\ \BBA {} Echikson, B.%
\end{APACrefauthors}%
\unskip\
\newblock
\APACrefYearMonthDay{2023}{}{}.
\newblock
\APACrefbtitle {{M}iddle {E}ast {V}iolence {T}ests {E}urope's {N}ew {D}igital
  {C}ontent {L}aw.} {{M}iddle {E}ast {V}iolence {T}ests {E}urope's {N}ew
  {D}igital {C}ontent {L}aw.}
\newblock
\APACrefnote{{C}enter for {E}uropean {P}olicy {A}nalysis ({CEPA}). {A}vailable
  online in October:
  \url{https://cepa.org/article/mideast-violence-tests-europes-new-dsa-digital-content-law/}}
\PrintBackRefs{\CurrentBib}

\bibitem [\protect \citeauthoryear {%
Ruohonen%
}{%
Ruohonen%
}{%
{\protect \APACyear {2023}}%
{\protect \APACexlab {{\protect \BCnt {1}}}}}]{%
Ruohonen23IPR}
\APACinsertmetastar {%
Ruohonen23IPR}%
\begin{APACrefauthors}%
Ruohonen, J.%
\end{APACrefauthors}%
\unskip\
\newblock
\APACrefYearMonthDay{2023{\protect \BCnt {1}}}{}{}.
\newblock
\APACrefbtitle {{A} {N}ote on the {P}roposed {L}aw for {I}mproving the
  {T}ransparency of {P}olitical {A}dvertising in the {E}uropean {U}nion.} {{A}
  {N}ote on the {P}roposed {L}aw for {I}mproving the {T}ransparency of
  {P}olitical {A}dvertising in the {E}uropean {U}nion.}
\newblock
\APACrefnote{{A}rchvied manuscript, available online:
  \href{https://arxiv.org/abs/2303.02863}{arXiv:2303.02863}}
\PrintBackRefs{\CurrentBib}

\bibitem [\protect \citeauthoryear {%
Ruohonen%
}{%
Ruohonen%
}{%
{\protect \APACyear {2023}}%
{\protect \APACexlab {{\protect \BCnt {2}}}}}]{%
Ruohonen23CLSR}
\APACinsertmetastar {%
Ruohonen23CLSR}%
\begin{APACrefauthors}%
Ruohonen, J.%
\end{APACrefauthors}%
\unskip\
\newblock
\APACrefYearMonthDay{2023{\protect \BCnt {2}}}{}{}.
\newblock
\APACrefbtitle {{A} {T}ext {M}ining {A}nalysis of {D}ata {P}rotection
  {P}olitics: {T}he {C}ase of {P}lenary {S}essions of the {E}uropean
  {P}arliament.} {{A} {T}ext {M}ining {A}nalysis of {D}ata {P}rotection
  {P}olitics: {T}he {C}ase of {P}lenary {S}essions of the {E}uropean
  {P}arliament.}
\newblock
\APACrefnote{{A}rchived manuscript. Available online:
  \url{https://arxiv.org/abs/2302.09939}}
\PrintBackRefs{\CurrentBib}

\bibitem [\protect \citeauthoryear {%
Ruohonen%
\ \BBA {} Hjerppe%
}{%
Ruohonen%
\ \BBA {} Hjerppe%
}{%
{\protect \APACyear {2022}}%
}]{%
Ruohonen22IS}
\APACinsertmetastar {%
Ruohonen22IS}%
\begin{APACrefauthors}%
Ruohonen, J.%
\BCBT {}\ \BBA {} Hjerppe, K.%
\end{APACrefauthors}%
\unskip\
\newblock
\APACrefYearMonthDay{2022}{}{}.
\newblock
{\BBOQ}\APACrefatitle {{T}he {GDPR} {E}nforcement {F}ines at {G}lance} {{T}he
  {GDPR} {E}nforcement {F}ines at {G}lance}.{\BBCQ}
\newblock
\APACjournalVolNumPages{Information Systems}{106}{}{101876}.
\newblock

\newblock

\PrintBackRefs{\CurrentBib}

\bibitem [\protect \citeauthoryear {%
Ruohonen%
, Hyrynsalmi%
\BCBL {}\ \BBA {} Lepp\"anen%
}{%
Ruohonen%
\ \protect \BOthers {.}}{%
{\protect \APACyear {2020}}%
}]{%
Ruohonen20CHB}
\APACinsertmetastar {%
Ruohonen20CHB}%
\begin{APACrefauthors}%
Ruohonen, J.%
, Hyrynsalmi, S.%
\BCBL {} Lepp\"anen, V.%
\end{APACrefauthors}%
\unskip\
\newblock
\APACrefYearMonthDay{2020}{}{}.
\newblock
{\BBOQ}\APACrefatitle {{A} {M}ixed {M}ethods {P}robe into the {D}irect
  {D}isclosure of {S}oftware {V}ulnerabilities} {{A} {M}ixed {M}ethods {P}robe
  into the {D}irect {D}isclosure of {S}oftware {V}ulnerabilities}.{\BBCQ}
\newblock
\APACjournalVolNumPages{Computers in Human Behavior}{103}{}{161--173}.
\newblock

\newblock

\PrintBackRefs{\CurrentBib}

\bibitem [\protect \citeauthoryear {%
Ruohonen%
\ \BBA {} Lepp\"anen%
}{%
Ruohonen%
\ \BBA {} Lepp\"anen%
}{%
{\protect \APACyear {2017}}%
}]{%
Ruohonen17EISIC}
\APACinsertmetastar {%
Ruohonen17EISIC}%
\begin{APACrefauthors}%
Ruohonen, J.%
\BCBT {}\ \BBA {} Lepp\"anen, V.%
\end{APACrefauthors}%
\unskip\
\newblock
\APACrefYearMonthDay{2017}{}{}.
\newblock
{\BBOQ}\APACrefatitle {{W}hose {H}ands {A}re in the {F}innish {C}ookie {J}ar?}
  {{W}hose {H}ands {A}re in the {F}innish {C}ookie {J}ar?}{\BBCQ}
\newblock
 \APACrefbtitle {{P}roceedings of the {E}uropean {I}ntelligence and {S}ecurity
  {I}nformatics Conference ({EISIC} 2017)} {{P}roceedings of the {E}uropean
  {I}ntelligence and {S}ecurity {I}nformatics conference ({EISIC} 2017)}\
  (\BPGS\ 127--130).
\newblock
\APACaddressPublisher{Athens}{IEEE}.
\PrintBackRefs{\CurrentBib}

\bibitem [\protect \citeauthoryear {%
Ruohonen%
\ \BBA {} Lepp\"anen%
}{%
Ruohonen%
\ \BBA {} Lepp\"anen%
}{%
{\protect \APACyear {2018}}%
}]{%
Ruohonen18WPES}
\APACinsertmetastar {%
Ruohonen18WPES}%
\begin{APACrefauthors}%
Ruohonen, J.%
\BCBT {}\ \BBA {} Lepp\"anen, V.%
\end{APACrefauthors}%
\unskip\
\newblock
\APACrefYearMonthDay{2018}{}{}.
\newblock
{\BBOQ}\APACrefatitle {{I}nvisible {P}ixels {A}re {D}ead, {L}ong {L}ive
  {I}nvisible {P}ixels!} {{I}nvisible {P}ixels {A}re {D}ead, {L}ong {L}ive
  {I}nvisible {P}ixels!}{\BBCQ}
\newblock
 \APACrefbtitle {{P}roceedings of the 17th {W}orkshop on {P}rivacy in the
  {E}lectronic {S}ociety ({WPES} 2018)} {{P}roceedings of the 17th {W}orkshop
  on {P}rivacy in the {E}lectronic {S}ociety ({WPES} 2018)}\ (\BPGS\ 28--32).
\newblock
\APACaddressPublisher{Toronto}{ACM}.
\PrintBackRefs{\CurrentBib}

\bibitem [\protect \citeauthoryear {%
Ruohonen%
, Salovaara%
\BCBL {}\ \BBA {} Lepp\"anen%
}{%
Ruohonen%
\ \protect \BOthers {.}}{%
{\protect \APACyear {2018}}%
}]{%
Ruohonen18PST}
\APACinsertmetastar {%
Ruohonen18PST}%
\begin{APACrefauthors}%
Ruohonen, J.%
, Salovaara, J.%
\BCBL {} Lepp\"anen, V.%
\end{APACrefauthors}%
\unskip\
\newblock
\APACrefYearMonthDay{2018}{}{}.
\newblock
{\BBOQ}\APACrefatitle {{C}rossing {C}ross-{D}omain {P}aths in the {C}urrent
  {W}eb} {{C}rossing {C}ross-{D}omain {P}aths in the {C}urrent {W}eb}.{\BBCQ}
\newblock
 \APACrefbtitle {{P}roceedings of the 16th {A}nnual {C}onference on {P}rivacy,
  {S}ecurity and {T}rust ({PST} 2018)} {{P}roceedings of the 16th {A}nnual
  {C}onference on {P}rivacy, {S}ecurity and {T}rust ({PST} 2018)}\ (\BPGS\
  1--5).
\newblock
\APACaddressPublisher{Belfast}{IEEE}.
\PrintBackRefs{\CurrentBib}

\bibitem [\protect \citeauthoryear {%
S{\ae}tra%
, Coeckelbergh%
\BCBL {}\ \BBA {} Danaher%
}{%
S{\ae}tra%
\ \protect \BOthers {.}}{%
{\protect \APACyear {2022}}%
}]{%
Saetra22}
\APACinsertmetastar {%
Saetra22}%
\begin{APACrefauthors}%
S{\ae}tra, H.S.%
, Coeckelbergh, M.%
\BCBL {} Danaher, J.%
\end{APACrefauthors}%
\unskip\
\newblock
\APACrefYearMonthDay{2022}{}{}.
\newblock
{\BBOQ}\APACrefatitle {{T}he {AI} {E}thicist's {D}ilemma: {F}ighting {B}ig
  {T}ech by {S}upporting {B}ig {T}ech} {{T}he {AI} {E}thicist's {D}ilemma:
  {F}ighting {B}ig {T}ech by {S}upporting {B}ig {T}ech}.{\BBCQ}
\newblock
\APACjournalVolNumPages{AI Ethics}{2}{}{15--27}.
\newblock

\newblock

\PrintBackRefs{\CurrentBib}

\bibitem [\protect \citeauthoryear {%
Schlag%
}{%
Schlag%
}{%
{\protect \APACyear {2023}}%
}]{%
Schlag23}
\APACinsertmetastar {%
Schlag23}%
\begin{APACrefauthors}%
Schlag, G.%
\end{APACrefauthors}%
\unskip\
\newblock
\APACrefYearMonthDay{2023}{}{}.
\newblock
{\BBOQ}\APACrefatitle {{E}uropean {U}nion's {R}egulating of {S}ocial {M}edia:
  {A} {D}iscourse {A}nalysis of the {D}igital {S}ervices {A}ct} {{E}uropean
  {U}nion's {R}egulating of {S}ocial {M}edia: {A} {D}iscourse {A}nalysis of the
  {D}igital {S}ervices {A}ct}.{\BBCQ}
\newblock
\APACjournalVolNumPages{Politics and Governance}{11}{3}{168--177}.
\newblock

\newblock

\PrintBackRefs{\CurrentBib}

\bibitem [\protect \citeauthoryear {%
Shang%
, Feng%
\BCBL {}\ \BBA {} Chirag%
}{%
Shang%
\ \protect \BOthers {.}}{%
{\protect \APACyear {2022}}%
}]{%
Shang22}
\APACinsertmetastar {%
Shang22}%
\begin{APACrefauthors}%
Shang, R.%
, Feng, K.J.K.%
\BCBL {} Chirag, S.%
\end{APACrefauthors}%
\unskip\
\newblock
\APACrefYearMonthDay{2022}{}{}.
\newblock
{\BBOQ}\APACrefatitle {{W}hy {A}m {I} {N}ot {S}eeing {I}t? {U}nderstanding
  {U}sers' {N}eeds for {C}ounterfactual {E}xplanations in {E}veryday
  {R}ecommendations} {{W}hy {A}m {I} {N}ot {S}eeing {I}t? {U}nderstanding
  {U}sers' {N}eeds for {C}ounterfactual {E}xplanations in {E}veryday
  {R}ecommendations}.{\BBCQ}
\newblock
 \APACrefbtitle {{P}roceedings of the {C}onference on {F}airness,
  {A}ccountability, and {T}ransparency ({FAccT} 2022)} {{P}roceedings of the
  {C}onference on {F}airness, {A}ccountability, and {T}ransparency ({FAccT}
  2022)}\ (\BPGS\ 1330--1340).
\newblock
\APACaddressPublisher{Seoul}{ACM}.
\PrintBackRefs{\CurrentBib}

\bibitem [\protect \citeauthoryear {%
Shin%
, Rasul%
\BCBL {}\ \BBA {} Fotiadis%
}{%
Shin%
, Rasul%
\BCBL {}\ \BBA {} Fotiadis%
}{%
{\protect \APACyear {2022}}%
}]{%
Shin22b}
\APACinsertmetastar {%
Shin22b}%
\begin{APACrefauthors}%
Shin, D.%
, Rasul, A.%
\BCBL {} Fotiadis, A.%
\end{APACrefauthors}%
\unskip\
\newblock
\APACrefYearMonthDay{2022}{}{}.
\newblock
{\BBOQ}\APACrefatitle {{W}hy {A}m {I} {S}eeing {T}his? {D}econstructing
  {A}lgorithm {L}iteracy {T}hrough the {L}ens of {U}sers} {{W}hy {A}m {I}
  {S}eeing {T}his? {D}econstructing {A}lgorithm {L}iteracy {T}hrough the {L}ens
  of {U}sers}.{\BBCQ}
\newblock
\APACjournalVolNumPages{Internet Research}{32}{4}{1214--1234}.
\newblock

\newblock

\PrintBackRefs{\CurrentBib}

\bibitem [\protect \citeauthoryear {%
Shin%
, Zaid%
, Biocca%
\BCBL {}\ \BBA {} Rasul%
}{%
Shin%
, Zaid%
\BCBL {}\ \protect \BOthers {.}}{%
{\protect \APACyear {2022}}%
}]{%
Shin22a}
\APACinsertmetastar {%
Shin22a}%
\begin{APACrefauthors}%
Shin, D.%
, Zaid, B.%
, Biocca, F.%
\BCBL {} Rasul, A.%
\end{APACrefauthors}%
\unskip\
\newblock
\APACrefYearMonthDay{2022}{}{}.
\newblock
{\BBOQ}\APACrefatitle {{I}n {P}latforms {W}e {T}rust? {U}nlocking the
  {B}lack-{B}ox of {N}ews {A}lgorithms {T}hrough {I}nterpretable {AI}} {{I}n
  {P}latforms {W}e {T}rust? {U}nlocking the {B}lack-{B}ox of {N}ews
  {A}lgorithms {T}hrough {I}nterpretable {AI}}.{\BBCQ}
\newblock
\APACjournalVolNumPages{Journal of Broadcasting \& Electronic
  Media}{66}{2}{235--256}.
\newblock

\newblock

\PrintBackRefs{\CurrentBib}

\bibitem [\protect \citeauthoryear {%
Smith%
, Jayne%
\BCBL {}\ \BBA {} Burke%
}{%
Smith%
\ \protect \BOthers {.}}{%
{\protect \APACyear {2022}}%
}]{%
Smith22}
\APACinsertmetastar {%
Smith22}%
\begin{APACrefauthors}%
Smith, J.J.%
, Jayne, L.%
\BCBL {} Burke, R.%
\end{APACrefauthors}%
\unskip\
\newblock
\APACrefYearMonthDay{2022}{}{}.
\newblock
{\BBOQ}\APACrefatitle {{R}ecommender {S}ystems and {A}lgorithmic {H}ate}
  {{R}ecommender {S}ystems and {A}lgorithmic {H}ate}.{\BBCQ}
\newblock
 \APACrefbtitle {{P}roceedings of the 16th {ACM} {C}onference on {R}ecommender
  {S}ystems ({R}ec{S}ys 2022)} {{P}roceedings of the 16th {ACM} {C}onference on
  {R}ecommender {S}ystems ({R}ec{S}ys 2022)}\ (\BPGS\ 592--597).
\newblock
\APACaddressPublisher{Seattle}{ACM}.
\PrintBackRefs{\CurrentBib}

\bibitem [\protect \citeauthoryear {%
Srivastava%
}{%
Srivastava%
}{%
{\protect \APACyear {2021}}%
}]{%
Srivastava21}
\APACinsertmetastar {%
Srivastava21}%
\begin{APACrefauthors}%
Srivastava, S.%
\end{APACrefauthors}%
\unskip\
\newblock
\APACrefYearMonthDay{2021}{}{}.
\newblock
{\BBOQ}\APACrefatitle {{A}lgorithmic {G}overnance and the {I}nternational
  {P}olitics of {B}ig {T}ech} {{A}lgorithmic {G}overnance and the
  {I}nternational {P}olitics of {B}ig {T}ech}.{\BBCQ}
\newblock
\APACjournalVolNumPages{Perspectives on Politics}{}{}{1--12}.
\newblock

\newblock

\PrintBackRefs{\CurrentBib}

\bibitem [\protect \citeauthoryear {%
Starke%
, Baleis%
, Keller%
\BCBL {}\ \BBA {} Marcinkowski%
}{%
Starke%
\ \protect \BOthers {.}}{%
{\protect \APACyear {2022}}%
}]{%
Starke22}
\APACinsertmetastar {%
Starke22}%
\begin{APACrefauthors}%
Starke, C.%
, Baleis, J.%
, Keller, B.%
\BCBL {} Marcinkowski, F.%
\end{APACrefauthors}%
\unskip\
\newblock
\APACrefYearMonthDay{2022}{}{}.
\newblock
{\BBOQ}\APACrefatitle {{F}airness {P}erceptions of {A}lgorithmic
  {D}ecision-{M}aking: {A} {S}ystematic {R}eview of the {E}mpirical
  {L}iterature} {{F}airness {P}erceptions of {A}lgorithmic {D}ecision-{M}aking:
  {A} {S}ystematic {R}eview of the {E}mpirical {L}iterature}.{\BBCQ}
\newblock
\APACjournalVolNumPages{Big Data \& Society}{9}{2}{20539517221115189}.
\newblock

\newblock

\PrintBackRefs{\CurrentBib}

\bibitem [\protect \citeauthoryear {%
Strowel%
\ \BBA {} {De Meyere}%
}{%
Strowel%
\ \BBA {} {De Meyere}%
}{%
{\protect \APACyear {2023}}%
}]{%
Strowel23}
\APACinsertmetastar {%
Strowel23}%
\begin{APACrefauthors}%
Strowel, A.%
\BCBT {}\ \BBA {} {De Meyere}, J.%
\end{APACrefauthors}%
\unskip\
\newblock
\APACrefYearMonthDay{2023}{}{}.
\newblock
{\BBOQ}\APACrefatitle {{T}he {D}igital {S}ervices {A}ct: {T}ransparency as an
  {E}fficient {T}ool to {C}urb the {S}pread of {D}isinformation on {O}nline
  {P}latforms?} {{T}he {D}igital {S}ervices {A}ct: {T}ransparency as an
  {E}fficient {T}ool to {C}urb the {S}pread of {D}isinformation on {O}nline
  {P}latforms?}{\BBCQ}
\newblock
\APACjournalVolNumPages{Journal of Intellectual Property, Information
  Technology and Electronic Commerce Law}{14}{1}{66--83}.
\newblock

\newblock

\PrintBackRefs{\CurrentBib}

\bibitem [\protect \citeauthoryear {%
Tabassum%
\ \protect \BOthers {.}}{%
Tabassum%
\ \protect \BOthers {.}}{%
{\protect \APACyear {2020}}%
}]{%
Tabassum20}
\APACinsertmetastar {%
Tabassum20}%
\begin{APACrefauthors}%
Tabassum, M.%
, Kosi\'{n}ski, T.%
, Frik, A.%
, Malkin, N.%
, Wijesekera, P.%
, Egelman, S.%
\BCBL {} Lipford, H.R.%
\end{APACrefauthors}%
\unskip\
\newblock
\APACrefYearMonthDay{2020}{}{}.
\newblock
{\BBOQ}\APACrefatitle {{I}nvestigating {U}sers' {P}references and
  {E}xpectations for {A}lways-{L}istening {V}oice {A}ssistants}
  {{I}nvestigating {U}sers' {P}references and {E}xpectations for
  {A}lways-{L}istening {V}oice {A}ssistants}.{\BBCQ}
\newblock
\APACjournalVolNumPages{Proceedings of the ACM on Interactive, Mobile, Wearable
  and Ubiquitous Technologies}{3}{4}{1--23}.
\newblock

\newblock

\PrintBackRefs{\CurrentBib}

\bibitem [\protect \citeauthoryear {%
Teppan%
\ \BBA {} Zanker%
}{%
Teppan%
\ \BBA {} Zanker%
}{%
{\protect \APACyear {2015}}%
}]{%
Teppan15}
\APACinsertmetastar {%
Teppan15}%
\begin{APACrefauthors}%
Teppan, E.C.%
\BCBT {}\ \BBA {} Zanker, M.%
\end{APACrefauthors}%
\unskip\
\newblock
\APACrefYearMonthDay{2015}{}{}.
\newblock
{\BBOQ}\APACrefatitle {{D}ecision {B}iases in {R}ecommender {S}ystems}
  {{D}ecision {B}iases in {R}ecommender {S}ystems}.{\BBCQ}
\newblock
\APACjournalVolNumPages{Journal of Internet Commerce}{14}{2}{255--275}.
\newblock

\newblock

\PrintBackRefs{\CurrentBib}

\bibitem [\protect \citeauthoryear {%
Terren%
\ \BBA {} Borge%
}{%
Terren%
\ \BBA {} Borge%
}{%
{\protect \APACyear {2021}}%
}]{%
Terren21}
\APACinsertmetastar {%
Terren21}%
\begin{APACrefauthors}%
Terren, L.%
\BCBT {}\ \BBA {} Borge, R.%
\end{APACrefauthors}%
\unskip\
\newblock
\APACrefYearMonthDay{2021}{}{}.
\newblock
{\BBOQ}\APACrefatitle {{E}cho {C}hambers on {S}ocial {M}edia: {A} {S}ystematic
  {R}eview of the {L}iterature} {{E}cho {C}hambers on {S}ocial {M}edia: {A}
  {S}ystematic {R}eview of the {L}iterature}.{\BBCQ}
\newblock
\APACjournalVolNumPages{Review of Communication Research}{9}{}{99--118}.
\newblock

\newblock

\PrintBackRefs{\CurrentBib}

\bibitem [\protect \citeauthoryear {%
Tintarev%
\ \BBA {} Masthoff%
}{%
Tintarev%
\ \BBA {} Masthoff%
}{%
{\protect \APACyear {2007}}%
}]{%
Tintarev07}
\APACinsertmetastar {%
Tintarev07}%
\begin{APACrefauthors}%
Tintarev, N.%
\BCBT {}\ \BBA {} Masthoff, J.%
\end{APACrefauthors}%
\unskip\
\newblock
\APACrefYearMonthDay{2007}{}{}.
\newblock
{\BBOQ}\APACrefatitle {{A} {S}urvey of {E}xplanations in {R}ecommender
  {S}ystems} {{A} {S}urvey of {E}xplanations in {R}ecommender
  {S}ystems}.{\BBCQ}
\newblock
 \APACrefbtitle {{P}roceedings of the 23rd {I}nternational {C}onference on
  {D}ata {E}ngineering {W}orkshop ({CDE Wkshp} 2007)} {{P}roceedings of the
  23rd {I}nternational {C}onference on {D}ata {E}ngineering {W}orkshop ({CDE
  Wkshp} 2007)}\ (\BPGS\ 801--810).
\newblock
\APACaddressPublisher{Istanbul}{IEEE}.
\PrintBackRefs{\CurrentBib}

\bibitem [\protect \citeauthoryear {%
Tontodimamma%
, Nissi%
, Sarra%
\BCBL {}\ \BBA {} Fontanella%
}{%
Tontodimamma%
\ \protect \BOthers {.}}{%
{\protect \APACyear {2021}}%
}]{%
Tontodimmamma21}
\APACinsertmetastar {%
Tontodimmamma21}%
\begin{APACrefauthors}%
Tontodimamma, A.%
, Nissi, E.%
, Sarra, A.%
\BCBL {} Fontanella, L.%
\end{APACrefauthors}%
\unskip\
\newblock
\APACrefYearMonthDay{2021}{}{}.
\newblock
{\BBOQ}\APACrefatitle {{T}hirty {Y}ears of {R}esearch {I}nto {H}ate {S}peech:
  {T}opics of {I}nterest and {T}heir {E}volution} {{T}hirty {Y}ears of
  {R}esearch {I}nto {H}ate {S}peech: {T}opics of {I}nterest and {T}heir
  {E}volution}.{\BBCQ}
\newblock
\APACjournalVolNumPages{Scientometrics}{126}{}{157--179}.
\newblock

\newblock

\PrintBackRefs{\CurrentBib}

\bibitem [\protect \citeauthoryear {%
Treyger%
, Taylor%
, Kim%
\BCBL {}\ \BBA {} Holliday%
}{%
Treyger%
\ \protect \BOthers {.}}{%
{\protect \APACyear {2023}}%
}]{%
Treyger23}
\APACinsertmetastar {%
Treyger23}%
\begin{APACrefauthors}%
Treyger, E.%
, Taylor, J.%
, Kim, D.%
\BCBL {} Holliday, M.A.%
\end{APACrefauthors}%
\unskip\
\newblock
\APACrefYearMonthDay{2023}{}{}.
\newblock
\APACrefbtitle {{A}ssessing and {S}uing an {A}lgorithm: {P}erceptions of
  {A}lgorithmic {D}ecisionmaking.} {{A}ssessing and {S}uing an {A}lgorithm:
  {P}erceptions of {A}lgorithmic {D}ecisionmaking.}
\newblock
\APACrefnote{{RAND} {C}orporation. {A}vailable online in October:
  \url{https://www.rand.org/pubs/research_reports/RRA2100-1.html}}
\PrintBackRefs{\CurrentBib}

\bibitem [\protect \citeauthoryear {%
Tsai%
\ \BBA {} Brusilovsky%
}{%
Tsai%
\ \BBA {} Brusilovsky%
}{%
{\protect \APACyear {2021}}%
}]{%
Tsai21}
\APACinsertmetastar {%
Tsai21}%
\begin{APACrefauthors}%
Tsai, C.%
\BCBT {}\ \BBA {} Brusilovsky, P.%
\end{APACrefauthors}%
\unskip\
\newblock
\APACrefYearMonthDay{2021}{}{}.
\newblock
{\BBOQ}\APACrefatitle {{T}he {E}ffects of {C}ontrollability and
  {E}xplainability in a {S}ocial {R}ecommender {S}ystem} {{T}he {E}ffects of
  {C}ontrollability and {E}xplainability in a {S}ocial {R}ecommender
  {S}ystem}.{\BBCQ}
\newblock
\APACjournalVolNumPages{User Modeling and User-Adapted
  Interaction}{31}{}{591--627}.
\newblock

\newblock

\PrintBackRefs{\CurrentBib}

\bibitem [\protect \citeauthoryear {%
{TTP}%
}{%
{TTP}%
}{%
{\protect \APACyear {2023}}%
}]{%
TTP23}
\APACinsertmetastar {%
TTP23}%
\begin{APACrefauthors}%
{TTP}%
\end{APACrefauthors}%
\unskip\
\newblock
\APACrefYearMonthDay{2023}{}{}.
\newblock
\APACrefbtitle {{D}angerous by {D}esign: {Y}ou{T}ube {L}eads {Y}oung {G}amers
  to {V}ideos of {G}uns, {S}chool {S}hootings.} {{D}angerous by {D}esign:
  {Y}ou{T}ube {L}eads {Y}oung {G}amers to {V}ideos of {G}uns, {S}chool
  {S}hootings.}
\newblock
\APACrefnote{{T}ech {T}ransparency {P}roject ({TTP}). {A}vailable online in
  October:
  \url{https://www.techtransparencyproject.org/articles/youtube-leads-young-gamers-to-videos-of-guns-school}}
\PrintBackRefs{\CurrentBib}

\bibitem [\protect \citeauthoryear {%
Turillazzi%
, Taddeo%
, Floridi%
\BCBL {}\ \BBA {} Casolari%
}{%
Turillazzi%
\ \protect \BOthers {.}}{%
{\protect \APACyear {2023}}%
}]{%
Turillazzi23}
\APACinsertmetastar {%
Turillazzi23}%
\begin{APACrefauthors}%
Turillazzi, A.%
, Taddeo, M.%
, Floridi, L.%
\BCBL {} Casolari, F.%
\end{APACrefauthors}%
\unskip\
\newblock
\APACrefYearMonthDay{2023}{}{}.
\newblock
{\BBOQ}\APACrefatitle {{T}he {D}igital {S}ervices {A}ct: {A}n {A}nalysis of
  {I}ts {E}thical, {L}egal, and {S}ocial {I}mplications} {{T}he {D}igital
  {S}ervices {A}ct: {A}n {A}nalysis of {I}ts {E}thical, {L}egal, and {S}ocial
  {I}mplications}.{\BBCQ}
\newblock
\APACjournalVolNumPages{Law, Innovation and Technology}{15}{1}{83--106}.
\newblock

\newblock

\PrintBackRefs{\CurrentBib}

\bibitem [\protect \citeauthoryear {%
{UNESCO}%
}{%
{UNESCO}%
}{%
{\protect \APACyear {2023}}%
}]{%
UNESCO23}
\APACinsertmetastar {%
UNESCO23}%
\begin{APACrefauthors}%
{UNESCO}%
\end{APACrefauthors}%
\unskip\
\newblock
\APACrefYearMonthDay{2023}{}{}.
\newblock
\APACrefbtitle {{G}uidelines for the {G}overnance of {D}igital {P}latforms:
  {S}afeguarding {F}reedom of {E}xpression and {A}ccess to {I}nformation
  {T}hrough a {M}ulti-{S}takeholder {A}pproach.} {{G}uidelines for the
  {G}overnance of {D}igital {P}latforms: {S}afeguarding {F}reedom of
  {E}xpression and {A}ccess to {I}nformation {T}hrough a {M}ulti-{S}takeholder
  {A}pproach.}
\newblock
\APACrefnote{{T}he {U}nited {N}ations {E}ducational, {S}cientific and
  {C}ultural {O}rganization ({UNESCO}). {A}vailable online in November:
  \url{https://unesdoc.unesco.org/ark:/48223/pf0000387339}}
\PrintBackRefs{\CurrentBib}

\bibitem [\protect \citeauthoryear {%
{van Cleynenbreugel}%
}{%
{van Cleynenbreugel}%
}{%
{\protect \APACyear {2023}}%
}]{%
VanCleynenbreugel23}
\APACinsertmetastar {%
VanCleynenbreugel23}%
\begin{APACrefauthors}%
{van Cleynenbreugel}, P.%
\end{APACrefauthors}%
\unskip\
\newblock
\APACrefYearMonthDay{2023}{}{}.
\newblock
\APACrefbtitle {{D}igital {S}ervices {C}oordinators and {O}ther {C}ompetent
  {A}uthorities in the {D}igital {S}ervices {A}ct: {S}treamlined {E}nforcement
  {C}oordination {L}ost?} {{D}igital {S}ervices {C}oordinators and {O}ther
  {C}ompetent {A}uthorities in the {D}igital {S}ervices {A}ct: {S}treamlined
  {E}nforcement {C}oordination {L}ost?}
\newblock
\APACrefnote{{E}uropean {L}aw {B}log. {A}vailable online in December:
  \url{https://europeanlawblog.eu/2023/11/30/digital-services-coordinators-and-other-competent-authorities-in-the-digital-services-act-streamlined-enforcement-coordination-lost/}}
\PrintBackRefs{\CurrentBib}

\bibitem [\protect \citeauthoryear {%
{van den Broeck}%
, Poels%
\BCBL {}\ \BBA {} Walrave%
}{%
{van den Broeck}%
\ \protect \BOthers {.}}{%
{\protect \APACyear {2020}}%
}]{%
vanDenBroeck20}
\APACinsertmetastar {%
vanDenBroeck20}%
\begin{APACrefauthors}%
{van den Broeck}, E.%
, Poels, K.%
\BCBL {} Walrave, M.%
\end{APACrefauthors}%
\unskip\
\newblock
\APACrefYearMonthDay{2020}{}{}.
\newblock
{\BBOQ}\APACrefatitle {{H}ow {D}o {U}sers {E}valuate {P}ersonalized {F}acebook
  {A}dvertising? {A}n {A}nalysis of {C}onsumer- and {A}dvertiser {C}ontrolled
  {F}actors} {{H}ow {D}o {U}sers {E}valuate {P}ersonalized {F}acebook
  {A}dvertising? {A}n {A}nalysis of {C}onsumer- and {A}dvertiser {C}ontrolled
  {F}actors}.{\BBCQ}
\newblock
\APACjournalVolNumPages{Qualitative Market Research}{23}{2}{309--327}.
\newblock

\newblock

\PrintBackRefs{\CurrentBib}

\bibitem [\protect \citeauthoryear {%
{van Hoboken}%
}{%
{van Hoboken}%
}{%
{\protect \APACyear {2022}}%
}]{%
vanHoboken22}
\APACinsertmetastar {%
vanHoboken22}%
\begin{APACrefauthors}%
{van Hoboken}, J.%
\end{APACrefauthors}%
\unskip\
\newblock
\APACrefYearMonthDay{2022}{}{}.
\newblock
\APACrefbtitle {{E}uropean {L}essons in {S}elf-{E}xperimentation: {F}rom the
  {GDPR} to {E}uropean {P}latform {R}egulation.} {{E}uropean {L}essons in
  {S}elf-{E}xperimentation: {F}rom the {GDPR} to {E}uropean {P}latform
  {R}egulation.}
\newblock
\APACrefnote{{U}npublished {E}ssay, {C}entre for {I}nternational {G}overnance
  {I}nnovation {(CIGI)}. Available online in February 2023:
  \url{https://www.cigionline.org/articles/european-lessons-in-self-experimentation-from-the-gdpr-to-european-platform-regulation/}}
\PrintBackRefs{\CurrentBib}

\bibitem [\protect \citeauthoryear {%
Veale%
, Binns%
\BCBL {}\ \BBA {} Edwards%
}{%
Veale%
\ \protect \BOthers {.}}{%
{\protect \APACyear {2018}}%
}]{%
Veale18}
\APACinsertmetastar {%
Veale18}%
\begin{APACrefauthors}%
Veale, M.%
, Binns, R.%
\BCBL {} Edwards, L.%
\end{APACrefauthors}%
\unskip\
\newblock
\APACrefYearMonthDay{2018}{}{}.
\newblock
{\BBOQ}\APACrefatitle {{A}lgorithms {T}hat {R}emember: {M}odel {I}nversion
  {A}ttacks and {D}ata {P}rotection {L}aw} {{A}lgorithms {T}hat {R}emember:
  {M}odel {I}nversion {A}ttacks and {D}ata {P}rotection {L}aw}.{\BBCQ}
\newblock
\APACjournalVolNumPages{Philosophical Transactions of the Royal Society A:
  Mathematical, Physical and Engineering Sciences}{376}{2133}{1--15}.
\newblock

\newblock

\PrintBackRefs{\CurrentBib}

\bibitem [\protect \citeauthoryear {%
Waldman%
}{%
Waldman%
}{%
{\protect \APACyear {2021}}%
}]{%
Waldman21}
\APACinsertmetastar {%
Waldman21}%
\begin{APACrefauthors}%
Waldman, A.E.%
\end{APACrefauthors}%
\unskip\
\newblock
\APACrefYear{2021}.
\newblock
\APACrefbtitle {{I}ndustry {U}nbound: {T}he {I}nside {S}tory of {P}rivacy,
  {D}ata, and {C}orporate {P}ower} {{I}ndustry {U}nbound: {T}he {I}nside
  {S}tory of {P}rivacy, {D}ata, and {C}orporate {P}ower}.
\newblock
\APACaddressPublisher{Cambridge}{Cambridge University Press}.
\PrintBackRefs{\CurrentBib}

\bibitem [\protect \citeauthoryear {%
Whittaker%
, Looney%
, Reed%
\BCBL {}\ \BBA {} Votta%
}{%
Whittaker%
\ \protect \BOthers {.}}{%
{\protect \APACyear {2021}}%
}]{%
Whittaker21}
\APACinsertmetastar {%
Whittaker21}%
\begin{APACrefauthors}%
Whittaker, J.%
, Looney, S.%
, Reed, A.%
\BCBL {} Votta, F.%
\end{APACrefauthors}%
\unskip\
\newblock
\APACrefYearMonthDay{2021}{}{}.
\newblock
{\BBOQ}\APACrefatitle {{R}ecommender {S}ystems and the {A}mplification of
  {E}xtremist {C}ontent} {{R}ecommender {S}ystems and the {A}mplification of
  {E}xtremist {C}ontent}.{\BBCQ}
\newblock
\APACjournalVolNumPages{Internet Policy Review}{10}{2}{}.
\newblock

\newblock

\PrintBackRefs{\CurrentBib}

\bibitem [\protect \citeauthoryear {%
Yesilada%
\ \BBA {} Lewandowsky%
}{%
Yesilada%
\ \BBA {} Lewandowsky%
}{%
{\protect \APACyear {2022}}%
}]{%
Muhsin22}
\APACinsertmetastar {%
Muhsin22}%
\begin{APACrefauthors}%
Yesilada, M.%
\BCBT {}\ \BBA {} Lewandowsky, S.%
\end{APACrefauthors}%
\unskip\
\newblock
\APACrefYearMonthDay{2022}{}{}.
\newblock
{\BBOQ}\APACrefatitle {{S}ystematic {R}eview: {Y}ou{T}ube {R}ecommendations and
  {P}roblematic {C}ontent} {{S}ystematic {R}eview: {Y}ou{T}ube
  {R}ecommendations and {P}roblematic {C}ontent}.{\BBCQ}
\newblock
\APACjournalVolNumPages{Internet Policy Review}{11}{1}{1--22}.
\newblock

\newblock

\PrintBackRefs{\CurrentBib}

\bibitem [\protect \citeauthoryear {%
Zuboff%
}{%
Zuboff%
}{%
{\protect \APACyear {2020}}%
}]{%
Zuboff20a}
\APACinsertmetastar {%
Zuboff20a}%
\begin{APACrefauthors}%
Zuboff, S.%
\end{APACrefauthors}%
\unskip\
\newblock
\APACrefYearMonthDay{2020}{}{}.
\newblock
{\BBOQ}\APACrefatitle {{C}aveat {U}sor: {S}urveillance {C}apitalism as
  {E}pistemic {I}nequality} {{C}aveat {U}sor: {S}urveillance {C}apitalism as
  {E}pistemic {I}nequality}.{\BBCQ}
\newblock
 K.~Werbach\ (\BED), \APACrefbtitle {{A}fter the {D}igital {T}ornado:
  {N}etworks, {A}lgorithms, {H}umanity} {{A}fter the {D}igital {T}ornado:
  {N}etworks, {A}lgorithms, {H}umanity}\ (\BPGS\ 174--214).
\newblock
\APACaddressPublisher{Cambridge}{Cambridge University Press}.
\PrintBackRefs{\CurrentBib}

\end{thebibliography}

\end{document}